\documentclass[]{aastex631}

\submitjournal{ApJ}

\shorttitle{Biosignature Runaway}
\shortauthors{Ranjan et al.}
\graphicspath{{./}{figures/}}

\begin{document}

\title{Photochemical Runaway in Exoplanet Atmospheres: Implications for Biosignatures}

\author[0000-0002-5147-9053]{Sukrit Ranjan}
\affiliation{Northwestern University, Center for Interdisciplinary Exploration and Research in Astrophysics, Evanston, 60201, USA}
\affiliation{Northwestern University, Department of Physics \& Astronomy, Evanston, 60201, USA}
\affiliation{Blue Marble Space Institute of Science, Seattle, 98154, USA}
\affiliation{Massachusetts Institute of Technology, Department of Earth, Atmospheric \& Planetary Sciences, Cambridge, 02139, USA}

\author{Sara Seager}
\affiliation{Massachusetts Institute of Technology, Department of Earth, Atmospheric \& Planetary Sciences, Cambridge, 02139, USA}
\affiliation{Massachusetts Institute of Technology, Department of Physics, Cambridge, 02139, USA}
\affiliation{Massachusetts Institute of Technology, Department of Aeronautics \& Astronautics, Cambridge, 02139, USA}

\author{Zhuchang Zhan}
\affiliation{Massachusetts Institute of Technology, Department of Earth, Atmospheric \& Planetary Sciences, Cambridge, 02139, USA}

\author{Daniel D. B. Koll}
\affiliation{Massachusetts Institute of Technology, Department of Earth, Atmospheric \& Planetary Sciences, Cambridge, 02139, USA}

\author{William Bains}
\affiliation{Massachusetts Institute of Technology, Department of Earth, Atmospheric \& Planetary Sciences, Cambridge, 02139, USA}
\affiliation{School of Physics \& Astronomy, Cardiff University, 4 The Parade, Cardiff CF24 3AA, UK}

\author{Janusz J. Petkowski}
\affiliation{Massachusetts Institute of Technology, Department of Earth, Atmospheric \& Planetary Sciences, Cambridge, 02139, USA}

\author{Jingcheng Huang}
\affiliation{Massachusetts Institute of Technology, Department of Chemistry, Cambridge, 02139, USA}

\author[0000-0003-0525-9647]{Zifan Lin}
\affiliation{Massachusetts Institute of Technology, Department of Earth, Atmospheric \& Planetary Sciences, Cambridge, 02139, USA}




\begin{abstract}

About 2.5 billion years ago, microbes learned to harness plentiful solar energy to reduce CO$_2$ with H$_2$O, extracting energy and producing O$_2$ as waste. O$_2$ production from this metabolic process was so vigorous that it saturated its photochemical sinks, permitting it to reach ``runaway" conditions and rapidly accumulate in the atmosphere despite its reactivity. Here we argue that O$_2$ may not be unique: diverse gases produced by life may experience a ``runaway” effect similar to O$_2$. This runaway occurs because the ability of an atmosphere to photochemically cleanse itself of trace gases is generally finite. If produced at rates exceeding this finite limit, even reactive gases can rapidly accumulate to high concentrations and become potentially detectable. Planets orbiting smaller, cooler stars, such as the M dwarfs that are the prime targets for the James Webb Space Telescope (JWST), are especially favorable for runaway due to their lower UV emission compared to higher-mass stars. As an illustrative case study, we show that on a habitable exoplanet with an H$_2$-N$_2$ atmosphere and net surface production of NH$_3$ orbiting an M dwarf (the ``Cold Haber World" scenario), the reactive biogenic gas NH$_3$ can enter runaway, whereupon an increase in the surface production flux of one order of magnitude can increase NH$_3$ concentrations by three orders of magnitude and render it detectable by JWST in just two transits.  Our work on this and other gases suggests that diverse signs of life on exoplanets may be readily detectable at biochemically plausible production rates.

\end{abstract}

\keywords{Transmission spectroscopy (2133) --- Exoplanet atmospheric composition (2021) --- Extrasolar rocky planets (511) --- Habitable planets (695) --- Planetary atmospheres (1244) --- Biosignatures (2018)}


\section{Introduction}
The detection of biologically produced gases indicative of life ("biosignature gases") in rocky exoplanet atmospheres is a key goal of exoplanet science. Yet simulations predict that observations of even the main atmospheric gases will be challenging, mainly due to the small sizes of rocky planets and their atmospheres. Observations of trace biosignature gases will be even more challenging \citep{Morley2017, Batalha2018, Lustig-Yaeger2019}. On modern Earth, most products of biology do not accumulate to atmospheric concentrations that can be remotely detected over interstellar distances,  because either their production rates are low, or they are too photochemically reactive, or both  \citep{Kasting2014PNAS, Rugheimer2018, Kaltenegger2017}. Photochemical reactivity is also predicted to restrict diverse biosignature gases to undetectably low concentrations for exoplanets dissimilar to modern Earth as well, including gases for which efficient abiotic "false positive" production mechanisms have not yet been identified (e.g., \citealt{Domagal-Goldman2011, Sousa-Silva2020, Zhan2020}).

A notable exception to the pessimism regarding biosignature gases is oxygen (O$_2$). On modern Earth, O$_2$ constitutes 21\% of modern Earth's atmosphere by volume. Thanks to its high abundance, O$_2$ and its photochemical by-product O$_3$ are accessible to detection by next-generation telescopes for Earth-twin exoplanets orbiting small, nearby stars, though even these observations will be challenging \citep{Rodler2014, Reinhard2017, Kawashima2019, Fauchez2020}. While not unambiguous as a biosignature, the detectability of O$_2$ and its by-product O$_3$ make oxygen a prime target for biosignature searches \citep{Meadows2018}. However, the oxygenation of Earth's atmosphere was not straightforward. Earth's atmosphere was anoxic for approximately the first half of its history, and up to hundreds of millions of years may have elapsed between the emergence of biological O$_2$ production and its accumulation in the atmosphere to $>1$ ppmv levels. O$_2$'s accumulation to a dominant atmospheric gas ($>10\%$ by volume) is even more recent, within the last billion years  \citep{Lyons2014}. This is because O$_2$ is very reactive, and readily reacted with reductants in the atmosphere of early Earth. At low surface production fluxes, such reactions photochemically confine the atmospheric concentration of O$_2$ to undetectably low levels. However, when the net production flux of O$_2$ exceeded the supply of reductants to the atmosphere on a sustained basis, O$_2$ atmospheric concentrations rose dramatically, until ultimately limited by surface processes \citep{Goldblatt2006, Zahnle2006, Daines2017}.

O$_2$ may not be unique. At high enough production rates, diverse gases may come to saturate their photochemical sinks and increase rapidly in concentration as a function of surface flux. We term this general process "photochemical runaway," after the prototypical "CO runaway" \citep{Zahnle1986, Kasting2014}. The term “runaway” is often used to describe a time-dependent phenomenon, but despite its etymology, photochemical runaway refers to an essentially stationary effect\footnote{While primarily a steady-state, stationary effect, we note in passing there is an element of time dependence in runaway, because it may take time to saturate the surface, as occurred with O$_2$.}. The photochemical runaway of CO has been the most studied in the literature, but photochemical runaways of other gases (e.g., CH$_4$, PH$_3$, and isoprene) have also been briefly reported \citep{Kasting1990, Pavlov2003, Kharecha2005, Segura2005, Schwieterman2019, Sousa-Silva2020, Zhan2020}.  

The physical mechanism underlying photochemical runaway is that the photochemical sinks limiting trace gas accumulation in the atmosphere are finite, and can be overwhelmed with high gas production. More fully, photochemistry generally controls the removal of biogenic gases from habitable planet atmospheres, because thermochemistry is generally slow at habitable temperatures. Photochemical removal of atmospheric gases is mediated by UV photons, either directly, via photolysis, or indirectly, via reactions with radicals produced by photolysis \citep{CatlingKasting2017, Grenfell2018}. As the supply of UV photons and the supply of photolyzable substrates are both finite, the ability of the atmosphere to photochemically cleanse itself is also finite. If emitted at fluxes exceeding this finite threshold, gases can overwhelm the photochemical controls on their accumulation and rapidly increase in concentration, until they are ultimately limited by other processes, e.g. surface uptake by biogeochemical sinks. 

Photochemical runaway is important because it suggests that even reactive biosignature gases can accumulate to high, detectable concentrations at biochemically plausible fluxes, particularly for planets orbiting smaller, cooler stars with lower UV emission compared to higher-mass stars. Photochemical runaways are nonlinear: past the runaway threshold, a small increase in surface emission flux may result in dramatic increases in atmospheric concentration and hence detectability. The precise threshold for runaway varies by gas and planet scenario, but is often, though not always, set by the UV output of the host star. This is because it is UV photons that ultimately mediate the photochemical removal of atmospheric gases \citep{Segura2005, Zahnle2006}. Fortuitously for observations, this suggests that planets orbiting M-dwarf stars are more prone to runaways compared to planets orbiting Sun-like stars, because of their lower emission of photolytic near-UV (NUV) radiation \citep{Segura2005, Rugheimer2015mdwarf}. Older, cooler white dwarf stars should be similarly favorable for runaway due to their lower UV output (\citealt{Kozakis2020, Lin2022}). Due to the observational advantages derived from their small host stars, planets orbiting cool stars like M dwarfs and white dwarfs are the only plausible rocky planet targets for near-term atmospheric characterization via transmission spectroscopy. Thus, the planets most amenable to atmospheric characterization are also the most generally prone to runaways.

Photochemical runaway has been studied for individual gases as a photochemical phenomenon, but its observational implications, particularly for reactive biosignature gases, have not been made clear. Here, we study photochemical runaway as a general phenomenon, with emphasis on its observational implications. We illustrate the effect of photochemical runaway and its implications for trace gas detectability in exoplanet atmospheres by exploring the case study of NH$_3$ runaway in an H$_2$-N$_2$ exoplanet atmosphere. We discuss the generality of photochemical runaway, showing that with the same photochemical model and reaction network,  a broad range of gases in diverse planet scenarios, including non-H$_2$ dominated atmospheres, can undergo runaway. We address literature concerns regarding the physicality of runaway, demonstrating that the previous blanket dismissal of runaway on thermodynamic grounds is not justified and interpreting the rise of O$_2$ on Earth as an actualized example of photochemical runaway, giving empirical grounding to the theory. We close with a discussion of the observational implications and uncertainties of runaway.

\section{Methods\label{sec:methods}}

We studied NH$_3$ as a case study for photochemical runaway and its implications for the detectability of trace biosignature gases. To simulate NH$_3$ runaway, we choose an H$_2$-N$_2$-dominated atmospheric scenario, corresponding to the Cold Haber World scenario \citep{Seager2013, Seager2013b, Phillips2021} in which life would have a metabolic incentive for the high production of NH$_3$. We model the photochemical runaway of NH$_3$ in this planetary scenario (Section~\ref{sec:nh3_photochem_methods}), the implications of runaway for the detectability of NH$_3$ (Section~\ref{sec:seas}), and its climate feedback (Section~\ref{sec:climate_methods}). 

\subsection{Photochemical Modeling\label{sec:nh3_photochem_methods}}
To calculate the concentration of NH$_3$ as a function of surface emission flux, we employed the 1D photochemical model of \citet{Hu2012}. The model is designed for the flexible exploration of atmospheres of varying redox state, and is fully detailed in \citet{Hu2012, Hu2013}. In brief, the model calculates the composition of a rocky planet atmosphere in kinetic steady-state by solving the 1D continuity-transport equation. It encodes 111 CHNOS-bearing species linked by 900 chemical reactions, of which subsets can be flexibly specified to explore diverse atmospheric scenarios. The processes encoded by the model include surface emission, wet and dry deposition, eddy diffusion, molecular diffusion and diffusion-limited escape of H and H$_2$, formation and deposition of S$_8$ and H$_2$SO$_4$ aerosols, and photolysis; radiative transfer is computed via the delta two-stream approximation, including molecular absorption, Rayleigh scattering, and aerosol scattering and absorption. The model reproduces the trace gas compositions of the atmospheres of modern Earth and Mars, and has recently been intercompared with other models for the case of prebiotic Earth-analog exoplanets orbiting Sunlike stars\citep{Ranjan2020}. Relative to the original model, we have  corrected the $\geq 202$ nm CO$_2$ cross-sections, corrected the temperature dependence of the Henry's Law constants used in the wet deposition calculation, and updated the H$_2$O cross-sections to utilize the new measurements from \citet{Ranjan2020} (their "extrapolation" prescription). Prior to utilizing the updated model, we confirmed that the model reproduces the atmospheric composition of Earth and Mars, as defined in \citet{Hu2012}. 

\subsubsection{Photochemical Network}
We build on the chemical network of the \citet{Hu2012} atmospheric benchmark scenarios. We expand our reaction network relative to the \citet{Hu2012} benchmark scenarios by including nitrogenous chemistry. While the chemistry of nitrogen-bearing molecules was encoded into the Hu model, these reactions were not included in the reaction network used to study the \citet{Hu2012} abiotic exoplanet benchmark scenarios, because nitrogenous chemistry was not their focus. As our focus is nitrogenous NH$_3$, we instead include all nitrogenous chemistry encoded in the Hu model in our reaction network. We exclude reactions T45 and R92 of \citet{Hu2012} from our network, because they correspond to isomers of species in our model, and were originally included in error. Specifically, T45 refers to the thermal decay of HSOO and R92 refers to a reaction of CH$_2$=C \citep{Goumri1999, Laufer2004}; neither species is included in our model. We update the rate law for reaction R317 of \citet{Hu2012} (NH$_2$ + CH$_4$ $\rightarrow$ NH$_3$ + CH$_3$), from $8.77\times10^{-15} \times (\frac{T}{298K})^{3}\exp(\frac{-2130.0 K}{T})$ cm$^{3}$ s$^{-1}$
\citep{Moller1984} to $5.75\times10^{-11} \exp(\frac{-6951.7K}{T})$ cm$^{3}$ s$^{-1}$ \citep{Siddique2017}. Finally, we correct the rate law for reaction R40 of \citet{Hu2012} (NH + OH $\rightarrow$ NH$_2$ + O) from $2.94\times10^{-12}(\frac{T}{298.0K})^{0.1} \exp(\frac{5800.0K}{T})$ cm$^{3}$ s$^{-1}$ to $2.94\times10^{-12}(\frac{T}{298.0K})^{0.1} \exp(\frac{-5800.0K}{T})$ cm$^{3}$ s$^{-1}$ \citep{Cohen1991}. 

We follow \citet{Hu2012} in excluding reactions involving molecules containing $>2$ carbons. We exclude the $C>2$ chemistry because the photochemistry of the higher hydrocarbons and their accompanying haze is poorly understood; we account for it in an ad hoc fashion as in \citet{Hu2012}, i.e. by assigning a high deposition velocity of $1\times10^{-5}$ cm s$^{-1}$ to C$_2$H$_6$. This means that our model is invalid in hazy atmospheric regimes, e.g. when the CH$_4$/CO$_2$ ratio exceeds $\sim0.1$ in an N$_2$-dominated atmosphere; we do not deploy our model in this regime. In total, our network encompasses 736 reactions linking 86 chemical species. 

\subsubsection{Stellar Irradiation\label{sec:methods_stellarirrad}}
We consider irradiation from Sun-like stars and M dwarfs. We draw our irradiation spectrum for Sun-like stars from \citet{Hu2012}, who in turn synthesized it from the Air Mass Zero reference spectrum produced by the American Society For Testing and Materials\footnote{https://www.nrel.gov/grid/solar-resource/spectra.html} ($\geq119.5$ nm) and from the average quiet Sun emission spectrum of \citet{Curdt2004} ($<119.5$ nm). We represent M dwarf absorption spectra by that of GJ 876, and draw its spectrum from that synthesized by the MUSCLES\footnote{https://archive.stsci.edu/prepds/muscles/} collaboration based on telescopic measurements (v. 2.2; \citealt{France2016, Youngblood2016, Loyd2016}). We chose GJ 876 for its relatively well-characterized spectrum and for its low UV output relative to other M dwarfs with well-characterized spectra, which maximized the contrast with Sunlike stars to illustrate the impact of different UV irradiation fields. This low UV output means that calculations of biosignature gas buildup utilizing GJ 876 represent a favorable endmember scenario; this bias is counterbalanced by our expectation that the sample of M dwarfs with well-characterized UV radiation fields should be biased to stars with higher UV output, because these brighter stars are more amenable to observation. We follow \citet{Hu2012} in adopting a semimajor axis for the H$_2$-dominated atmosphere of 1.6 AU for the solar instellation case (based on crude climate calculations), and we scale our GJ 876 spectra to the same total instellation.

\subsubsection{Convergence Criteria}
We require that the chemical variation timescale \citep{Hu2012} of each species at each altitude with abundance exceeding 1 cm$^{-3}$ be $\geq10^{200}$ s. We adopt this unphysically stringent criteria, which corresponds to timescales far greater than the age of the universe, because we found that adopting less stringent criteria could sometimes lead to "false minima" that did not correspond to the true steady-state solution. For example, the O$_2$ instability as a function of boundary condition discussed in Section~\ref{sec:planet_scenarios} would not have been detected for a less stringent timescale requirement, e.g. one corresponding to the age of the solar system. For each simulation, we verify stable convergence by rerunning the converged solutions. 

We comment that it is particularly difficult to converge solutions according to these criteria when a gas enters runaway, particularly a photochemically reactive gas like NH$_3$. We attribute this to the observation that when a gas enters runaway, it begins to photochemically reengineer the atmosphere, e.g. CO suppressing OH; O$_2$ converting the atmosphere from near-neutral to strongly oxidizing (Appendix~\ref{sec:runaway_eg}). This makes the "runaway regime" photochemically unstable, as the bulk atmospheric forcing is changing. After runaway is complete and the atmosphere has been reengineered, convergence becomes easier. We therefore consider our calculations most precise for surface fluxes much above or below the runaway threshold, while solutions in close vicinity to the runaway threshold are more likely sensitive to physical and numerical uncertainties (e.g., uncertainties in the photochemical network, \citealt{Ranjan2020}). In other words, we consider the high gas concentrations at production fluxes significantly exceeding the runaway threshold and the low gas concentrations at production fluxes significantly below the runaway threshold to be robust, but do not place much confidence in the precise mapping between surface fluxes and atmospheric concentrations right at the runaway threshold.

\subsubsection{Planet Scenario\label{sec:planet_scenarios}}
We construct our planet scenario by modifying the H$_2$-dominated atmospheric benchmark scenario of \citet{Hu2012}. This scenario corresponds to an atmosphere with bulk composition of 0.9 bar H$_2$ and 0.1 bar N$_2$, overlying an abiotic but habitable planet, with a surface temperature of 288 K, an ocean, and Earth-like volcanic outgassing of H$_2$, CH$_4$, SO$_2$, and H$_2$S. The presence of an ocean is simulated by fixing the surface H$_2$O mixing ratio to be 0.01, corresponding to a fixed relative humidity of about 60\%. We adopt vertical temperature, pressure, and eddy diffusion profiles from \citet{Hu2012}, following logic detailed therein. We do not include the effects of lightning, which we would expect to produce reducing nitrogen species in an H$_2$-N$_2$ atmosphere; however, we expect such production to be modest, by analogy with NO production in more oxidizing atmospheres. Hence, our approach may slightly underestimate pNH$_3$. Table~\ref{tab:planet_param_nh3} summarizes the bulk parameters of the planetary scenario assumed in the photochemistry calculation. 

 \begin{deluxetable}{lr}
\tablecaption{Planetary Parameters Assumed in the Photochemistry Calculation of NH$_3$ Runaway\label{tab:planet_param_nh3}}
\tablewidth{0pt}
\tablehead{
\colhead{Parameter} & \colhead{Value} 
 }
\startdata
Bulk Composition & 90\% H$_2$, 10\% N$_2$\\
Surface Pressure (bar) & 1 \\
Surface Water Vapor Mixing Ratio & 0.01\\
Surface Temperature (K) & 288 \\
Stratospheric Temperature (K) & 160 \\
Stellar Irradiation Considered & Sun, GJ 876\\
Stellar Constant Relative to Modern Earth & 0.4\\
Planet Mass  ($M_{\Earth}$)& 1\\
Planet Radius ($R_{\Earth}$) & 1 \\
\enddata
\tablecomments{The bulk planetary properties correspond to the H$_2$-dominated (reducing) exoplanet atmospheric benchmark scenario of \citet{Hu2012}. Detailed chemical boundary conditions are presented in Table~\ref{tab:species_bcs}.} 
\end{deluxetable}

Photochemical models require the specification of chemical boundary conditions. Specifying such boundary conditions requires making assumptions regarding the geology and biology of the underlying planet; these assumptions are somewhat arbitrary. To achieve our specific goal of studying the photochemical limits to the runaway of NH$_3$, we choose a planet scenario that neglects biological activity, except for NH$_3$ production and loss. This choice enables a close comparison with the abiotic exoplanet benchmark scenario of \citet{Hu2012}, and hence isolation of the effects of the phenomenon of biosignature runaway. We neglect wet deposition of H$_2$, O$_2$, CO, CH$_4$, C$_2$H$_2$, C$_2$H$_4$, C$_2$H$_6$, and NH$_3$, under the assumption that on a planet with an otherwise inefficient biosphere the ocean would saturate in these compounds \citep{Hu2012}. A consequence of assuming a mostly abiotic planet is generally low deposition velocities in the absence of biological consumption. As a sensitivity test, we consider higher deposition velocities drawn from other works in the literature, and allow rainout of all species except NH$_3$ (i.e. no oceanic saturation); our conclusions are unaffected. Table~\ref{tab:species_bcs} gives both the low- and high-deposition boundary conditions considered in this work. 

Following the Cold Haber World scenario, we assume that the biosphere is a net source of NH$_3$ and sufficiently productive to saturate the surface of the planet in NH$_3$, as is the case for O$_2$ on Earth. We represent this in the simulation by turning off wet deposition of the biosignature gas and setting the dry deposition velocity to an $v_{dep}=10^{-7}$ cm s$^{-1}$. This $v_{dep}$ corresponds to the effective $v_{dep}$ of O$_2$ on modern Earth, scaled by $10\times$ for conservatism ($v_{dep}=10\times(\phi_{O_{2}}/n_{0{2}})=10\times((4\times10^{10}~\text{cm}^{-2}~\text{s}^{-1})/(0.21\times2.5\times10^{19} ~\text{cm}^{-3}))=10^{-7} ~\text{cm}\text{s}^{-1}$, where we have taken $\phi_{O_{2}}$ from \citealt{Zahnle2006}). In reality, surface processes may produce higher effective deposition velocities, and limit NH$_3$ to lower concentrations \citep{Huang2020}. We therefore emphasize that our calculations here specifically focus on the lifting of the atmospheric \emph{photochemical} barrier to biosignature gas accumulation, not the surficial barrier, though even the latter may be overcome if the surface can be saturated by sufficiently intense production, as occurred with O$_2$ on Earth (see Section~\ref{sec:discussion}).
 
We emphasize the enforcement of mass balance in our photochemical modeling. We do not approximate short-lived chemical species as being in photochemical equilibrium \citep{James2018}. We also minimize the use of surface mixing ratio boundary conditions in favor of surface flux boundary conditions. We employ fixed mixing ratio boundary conditions only for the bulk atmospheric components and for H$_2$O, and we employ surface flux boundary conditions for all other species, including NH$_3$. Surface mixing ratio boundary conditions explicitly violate mass balance, and can lead to solutions that are unstable when repeated with the equivalent surface flux boundary conditions. We tested this by reproducing the work of \citet{Zahnle2006} for O$_2$ buildup on Earth. When employing fixed mixing ratio boundary conditions for O$_2$ and CH$_4$, as they did, we reproduced their result that pO$_2$ was bivalued as a function of $\phi_{O_{2}}$. However, when we employed the equivalent surface flux boundary conditions for O$_2$ and ran our model to high numerical precision, the solutions collapsed to the low pO$_2$ branch. We observed similar instability with the ATMOS photochemical model \citep{Arney2016}, and similar behavior has recently been reported in the literature \citep{Gregory2021}. In the absence of observationally-derived boundary conditions, we consider surface emission and deposition to be the better independent variables compared to surface mixing ratios, because they are set by the properties of the underlying planet.

 
\subsection{Simulated Transmission Spectra and Observations\label{sec:seas}}

We assess the detection of NH$_3$ via transmission spectroscopy with simulated James Webb Space Telescope (JWST) observations. We computed the radiative transfer using the "Simulated Exoplanet Atmosphere Spectra" (SEAS) model from \citet{Zhan2020} and calculated the instrumental noise using Pandexo from \citet{Batalha2017}. We simulate transmission spectra for a hypothetical 1.5 $R_{\oplus}$, 5$M_{\oplus}$ super-Earth with an H$_2$-dominated atmosphere transiting an M dwarf-star similar to GJ876. We choose such a large planet because a larger planet is more likely to retain the H$_2$-dominated atmosphere assumed in the Cold Haber World scenario. The planet radius and mass we adopt are consistent with a rocky planet \citep{Rogers2015, Zeng2019}. Our photochemistry calculation was performed for an Earth-sized planet; we project to the super-Earth we consider here by holding constant the "molecular mixing ratios as a function of pressure, which, to first order, are invariant to changes in surface gravity" \citep{Sousa-Silva2020}. In other words, we assume that projecting to a super-Earth only affects the spectrum by affecting the planet radius, and by altering the scale height and total atmospheric column (due to different surface gravity).  

We use the output of the photochemical model discussed above (the molecular mixing ratio profile) and calculate the optical depth of each layer of the atmosphere \citep{Seager2013b, Zhan2020}. As stellar radiation beams penetrate along the limb paths of each layer, the absorption along each path is calculated as $A = n_{i,j}\sigma_{i,j}l_i$ where $A$ is absorption, $n$ is number density, $\sigma$ is absorption cross-section and $l$ is pathlength. The subscript $i$ denotes each layer that the stellar radiation beam penetrates, and $j$ denotes each molecule. The height of each layer is selected to be the scale height of the atmosphere. Next, we calculate the transmittance ($T$) of each beam using the Beer-Lambert Law. Then, we compute the total effective height $h$ of the atmosphere by multiplying the absorption ($A$ = 1-$T$) by the atmosphere's scale height. Finally, we represent the total attenuated flux as transit depth $(R_{planet} + h)^2/R_{star}^2$ in units of ppm.

We simulate the detection of the planet's atmosphere by calculating the signal-to-noise ratio (SNR) for JWST NIRSpec G395M \citep{Gardner2006}, which provided spectral coverage from 1 to 5 $\mu$m. Our simulations include randomly generated noise. We employ a two-part procedure. First, we determine the number of transits required to detect NH$_3$ at high confidence. To do so, we simulate observations of duration ranging from 1 to 100 transits, and identify the minimum number of transits at which NH$_3$ is detected at high confidence ($\gtrsim5\sigma$); here, two transits were required. Second, to ensure our results are not artifacts of a randomly low-noise simulation, we conduct 100 simulated two-transit observations, and report the average significance with which NH$_3$ is detected across these 100 simulations. The same photochemical solution is used for all simulated observations. In converting the integration time to the number of transits, we assume an impact parameter of 0, hence maximizing the time in transit and meaning that our calculations represent a best-case scenario for the number of transits required to detect the trace gas species ($T_{transit}= 3.2$ hours; Appendix~\ref{sec:ap_stellarparams}). We use the molecular absorption parameters from HITRAN \citep{Gordon2017}. We outline the details of the detection metric in \citet{Zhan2020}. 


\subsection{Climate Calculation \label{sec:climate_methods}}
For the purpose of photochemical sensitivity tests (Appendix~\ref{sec:runaway_nh3}), we estimate the climate impact of an NH$_3$ runaway in a H$_2$-N$_2$ atmosphere using a 1D climate model with line-by-line radiative transfer \citep{KollCronin2019}. The model assumes a moist-adiabatic troposphere at fixed relative humidity, capped by an isothermal stratosphere. For a given choice of surface temperature $T_{surf}$ and stratospheric temperature $T_{strat}$ we compute the top of atmosphere and stratospheric energy budgets, and iterate until both are in equilibrium. The model was previously validated in the runaway greenhouse limit, and for this work we additionally validated it against CH$_4$ warming calculations of early Mars \citep{Wordsworth2017}. The base climate is a pure H$_2$-N$_2$ atmosphere with $T_{surf}=288$ K and $T_{strat}=170$ K. 

\section{Results}

\begin{figure}[h]
\centering
\includegraphics[width=0.8\linewidth]{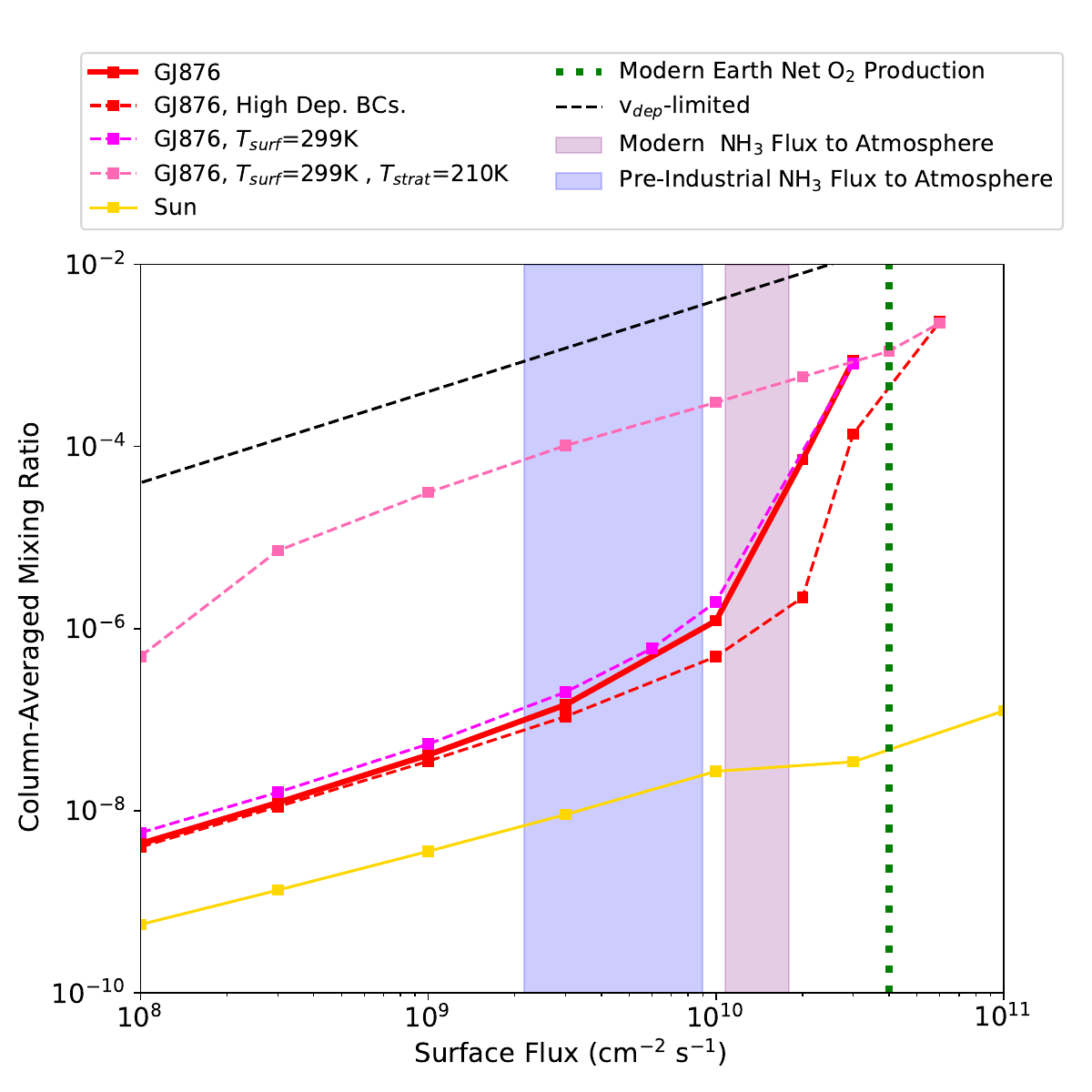}
\caption{NH$_3$ column-averaged mixing ratio as a function of net surface flux for an Earth-sized planet with an H$_2$-dominated atmosphere for the Cold Haber World scenario \citep{Seager2013, Seager2013b}. NH$_3$ surface deposition is assumed to be negligible due to surface saturation in this scenario. Our standard case (the red solid line) corresponds to a planet orbiting an M dwarf star with low wet and dry deposition of atmospheric species (Table~\ref{tab:species_bcs}; Section~\ref{sec:nh3_photochem_methods}), representing a planet with inefficient biological consumption of atmospheric species. We also show sensitivity test calculations for an otherwise identical planet with high wet and dry deposition of non-NH$_3$ species (the red dashed line; Table~\ref{tab:species_bcs}), with elevated surface temperature (the pink dashed line), and with elevated surface and stratospheric temperatures (the hot pink dashed line) due to assumed warming by NH$_3$ and/or its photochemical products. The solid yellow line shows NH$_3$ accumulation for a Sun-like stellar host. Modern biological O$_2$ production (net of biological consumption) is demarcated by a green line \citep{Zahnle2006}, and estimates of modern and preindustrial NH$_3$ flux to the atmosphere are represented by the purple and blue shaded regions, respectively \citep{Bouwman1997, Zhu2015}. NH$_3$ enters photochemical runaway at biochemically plausible surface production fluxes for M-dwarfs, but not Sun-like stars.}
\label{fig:solar_gj876_nh3_runaway}
\end{figure}

We hypothesize that photochemical runaway is a general phenomenon controlling the atmospheric abundance of gases in a planetary atmosphere. As a case study, we illustrate this phenomenon and its observational implications for exoplanets with an analysis of the runaway of NH$_3$ (ammonia) on a Cold Haber World with an H$_2$-N$_2$ atmosphere orbiting a small, cool M dwarf star (Figure~\ref{fig:solar_gj876_nh3_runaway}). Cold Haber Worlds are hypothetical planets where biological NH$_3$ production is intense enough to saturate the surface with NH$_3$ and suppress uptake by surface deposition, which would otherwise regulate it to undetectably low concentrations \citep{Seager2013, Seager2013b, Huang2020}. We choose to model the runaway of NH$_3$ because NH$_3$ is photochemically reactive (main photochemical loss mechanism: direct photolysis), and we wish to demonstrate that runaway can occur for even reactive gases, whose concentrations are otherwise undetectably low. We further choose NH$_3$ because NH$_3$ is a relatively well-studied gas whose photochemistry is comparatively well understood \citep{Kasting1982, Hu2012, Seager2013b, CatlingKasting2017}. We choose an H$_2$-N$_2$ atmosphere because life in such an atmosphere would have a metabolic incentive to produce excess NH$_3$, following the original scenario for the Cold Haber World. We choose an M dwarf host star because these small stars are the only class of stellar host around which near-term facilities can hope to characterize the atmospheres of small, rocky exoplanets, i.e. this is the observationally relevant case \citep{Cowan2015}. However, we argue that the phenomenon of runaway generalizes to all atmosphere types and stellar hosts, including non-H$_2$-dominated atmospheres and Sun-like stars, as shown in Appendix~\ref{sec:runaway_eg}, and by the example of terrestrial O$_2$.

We find NH$_3$ to enter runaway at biochemically plausible surface production fluxes for Cold Haber Worlds orbiting M dwarf stars (represented by GJ 876). Specifically, our simulations indicate NH$_3$ to enter runaway for $\phi_{NH_{3}}\geq 1\times10^{10}~\text{cm}^{-2}~\text{s}^{-1}$. For a production flux of $\phi_{NH_{3}}=2\times10^{10}~\text{cm}^{-2}~\text{s}^{-1}$, NH$_3$ concentrations are $\sim$ 70 ppmv. For comparison, the globally averaged modern NH$_3$ flux to Earth's atmosphere is estimated as $\phi_{NH_{3}}=1.1-1.8\times10^{10}~\text{cm}^{-2}~\text{s}^{-1}$ \citep{Bouwman1997}, and the preindustrial NH$_3$ flux is estimated as $\phi_{NH_{3}}=2-9\times10^{9}~\text{cm}^{-2}~\text{s}^{-1}$ ($2-5\times$ lower; \citealt{Zhu2015}), meaning that even a modern-Earth-like biosphere emits NH$_3$ to the atmosphere at rates within an order of magnitude of the runaway threshold for an M dwarf-hosted Cold Haber World. If NH$_3$ were net emitted at rates comparable to the net emission of O$_2$ on Earth ($\phi_{O_{2}}=4\times10^{10}~\text{cm}^{-2}~\text{s}^{-1}$ net, \citealt{Zahnle2006}), e.g., as a waste product of primary energy metabolism as in the Cold Haber World scenario,  then NH$_3$ would be in the runaway regime for M dwarf worlds. This statement is robust to sensitivity tests regarding climate and wet/dry deposition of non-NH$_3$ species. High surface deposition of non-NH$_3$ species increases the threshold for runaway by a factor of $2$, but does not prevent the phenomenon (Appendix~\ref{sec:runaway_nh3}). However, high surface deposition of NH$_3$ inhibits runaway: NH$_3$ runaway is critically dependent on the assumption that the surface is saturated in NH$_3$, as modern Earth's surface is saturated in O$_2$ (Section~\ref{sec:discussion}; \citealt{Huang2020}). Irradiation by a Sun-like UV field also suppresses runaway at the fluxes studied here, due to higher photolytic NUV irradiation. M dwarf planets are more amenable to runaway compared to G dwarf planets due to their lower NUV output, as shown previously for other gases like CH$_4$ \citep{Segura2005}. We find metabolic ammonia production to be thermodynamically profitable over the parameter space we model here (Appendix~\ref{sec:microbe-thermo}). Overall, the example of our biosphere suggests that NH$_3$ production fluxes compatible with photochemical runaway on M dwarf worlds are plausible, provided the critical assumption of surface saturation is met.

\begin{figure}[h]
\centering
\includegraphics[width=\linewidth]{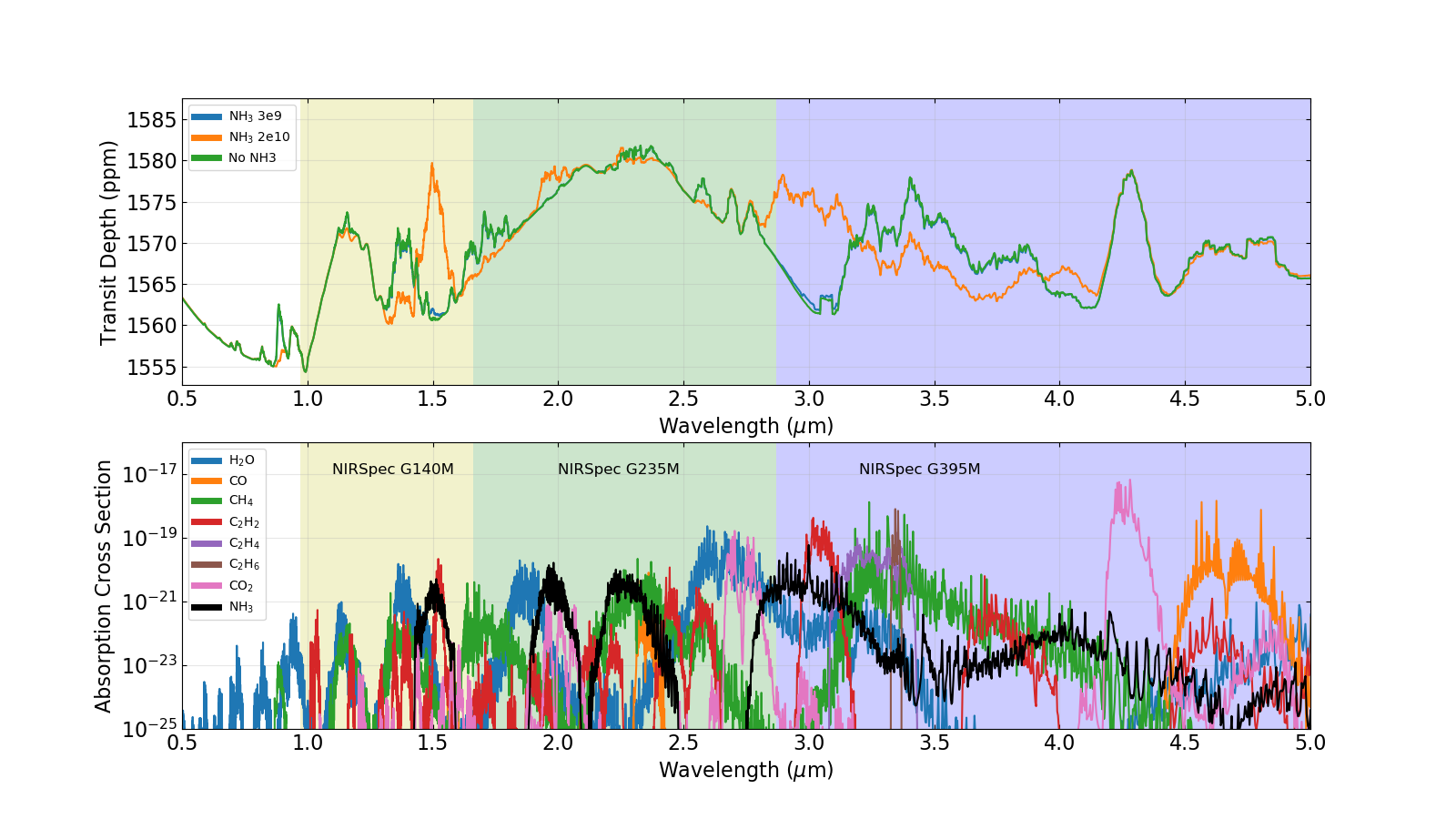}
\caption{The effect of photochemical runaway on the detectability of NH$_3$. Upper panel: simulated spectra of exoplanets with H$_2$-dominated atmospheres transiting an M dwarf star for NH$_3$ surface fluxes of 2$\times$ 10$^{10}~\text{cm}^{-2}~\text{s}^{-1}$ (orange),  3 $\times$ 10$^{9}~\text{cm}^{-2}~\text{s}^{-1}$ (blue), and  $0 \text{cm}^{-2}~\text{s}^{-1}$ (green). The y-axis shows transit depth (ppm), and the x-axis shows wavelength ($\mu$m). The spectra are simulated from 0.5 to 5 $\mu$m, covering the wavelength span of the JWST NIRSpec instrument. The yellow, green, and blue region shows the spectra coverage of NIRSpec 140 M, 235 M, and 395 M, respectively.  Lower panel: comparison of the NH$_3$ cross-sections with cross-sections of
dominant molecules (except H$_2$, whose spectral features are mainly dominated by collision-induced absorption) in the atmosphere such as H$_2$O, CO, CH$_4$, and CO$_2$. Photochemical runaway makes NH$_3$ detectable.}
\label{fig:nh3_spectra}
\end{figure}

The implications of photochemical runaway for NH$_3$ detectability are remarkable. In photochemical runaway (i.e. past a critical flux value), a mere order-of-magnitude increase in NH$_3$ production flux translates to the difference between NH$_3$ being essentially invisible and being highly detectable. We consider the detection of NH$_3$ on a Cold Haber World via transmission spectroscopy with the upcoming JWST, anticipated to be the premier near-term exoplanet characterization opportunity \citep{Cowan2015}. We do not expect NH$_3$ to be detectable at all on a 1.5 R$_{Earth}$ exoplanet with an H$_2$-dominated atmosphere orbiting a GJ 876-like star for a surface emission flux of $3\times10^{9}~\text{cm}^{-2}~\text{s}^{-1}$, below the runaway threshold. At this surface emission flux, NH$_3$ is  confined to near the surface and rapidly decays as a function of altitude, and the marginal absorption produced by NH$_3$ is not greater than the JWST noise floor. By comparison, 2 transits of JWST would suffice to detect the 1.45-1.55 $\mu$m feature of NH$_3$ with an average significance of $5.0\sigma$ for a surface emission flux of $2\times10^{10}~\text{cm}^{-2}~\text{s}^{-1}$, above the runaway threshold (Figure~\ref{fig:nh3_spectra}, ~\ref{fig:snr_nh3}; Appendix~\ref{sec:runaway_nh3}). At this surface emission flux, NH$_3$ is in runaway. In runaway, NH$_3$ concentrations are high and NH$_3$ populates the upper atmosphere, leading to significantly enhanced spectral features over the non-runaway cases.

The planet scenario we have considered corresponds to a rocky, habitable planet with an H$_2$-dominated atmosphere, for which life would have a strong metabolic incentive to produce NH$_3$. Such planets have no analog in the solar system, and it is not known whether they exist. Terrestrial-mass planets overlaid by H$_2$-dominated atmospheres are predicted for second-generation planets orbiting white-dwarf stars \citep{Lin2022} and are allowed by theory for main-sequence stars depending on extreme-UV irradiation level \citep{Owen2020}. Habitable (but not rocky) planets overlaid by H$_2$-dominated atmospheres have been proposed based on exoplanet mass/radius measurements \citep{Madhusudhan2021}, and H$_2$-dominated atmospheres are experimentally demonstrated to be compatible with life \citep{Seager2020}. However, the main motivation for considering planets with H$_2$-dominated atmospheres is observational: such planets are the only habitable worlds whose atmospheres will be generally accessible to spectral characterization over the next one to two decades, due to their light (therefore extended) atmospheres. Even if rocky planets with H$_2$-dominated atmospheres do not exist, planets with H$_2$-\textit{rich} atmospheres are thermodynamically capable of supporting Cold Haber Worlds. For Cold Haber Worlds that do not feature H$_2$-dominated atmospheres, runaway would still occur, but NH$_3$ detection would require next-generation telescopes with a noise floor lower than JWST, a host star smaller and closer than GJ876 like TRAPPIST-1, or both \citep{Huang2020}. 

We reiterate that our intent is not to singularly emphasize NH$_3$ as a biosignature gas, but rather to use it as a case study for illustrating photochemical runaway and its implications for the detectability of trace atmospheric gases. Indeed, NH$_3$ is just one of a steadily growing list of biogenic gases that can undergo runaway \citep{Schwieterman2019, Sousa-Silva2020, Zhan2020}. Because the phenomenon of runaway rests on the general observation that a gas's atmospheric sinks are finite, we expect a broad range of gases in diverse habitable planetary scenarios to undergo runaway if emitted at sufficiently high fluxes, as shown in Appendix~\ref{sec:runaway_eg}. Specific pathological cases in which runaway fails can be identified with detailed photochemical measurements and modeling (Appendix~\ref{sec:runaway_fail}). Abiotic production can also drive runaways (e.g., impacts may lead to CO runaway; \citep{Kasting1990}); therefore, abiotic false-positive analyses remain essential for biosignature gases, in or out of runaway. 

\section{Discussion\label{sec:discussion}}
Photochemical runaway lifts the photochemical control on biosignature gas accumulation. Photochemical processes like photolysis are predicted to limit the concentrations of most gaseous products of biology to undetectable levels on habitable planets \citep{Domagal-Goldman2011, Kasting2014PNAS, Sousa-Silva2020, Zhan2020}. However, in the runaway phase, this control is lifted, making possible much higher, potentially detectable concentrations of biosignature gases. For example, the biosignature gases PH$_3$ and isoprene are challenging-to-impossible to detect on habitable planets with JWST \textit{unless} they enter a runaway phase \citep{Sousa-Silva2020, Zhan2020}. 

Researchers disagree on whether photochemical runaways are physically realistic. While photochemical runaways arise in a range of models, some researchers postulate runaways to be unphysical, and set model boundary conditions that prevent them from occurring. For example, \citet{Segura2005} argue that CH$_4$ runaway on a modern Earth-like planet is implausible because the high pCH$_4$ and temperature associated with CH$_4$ runaway should thermodynamically inhibit methanogenesis, which other workers extend into a general dismissal of runaway \citep{Rugheimer2015mdwarf, Rugheimer2018}. However, this objection is specific to planets similar to the modern Earth. On planets analogous to early Earth (i.e. with high pCO$_2$ and pH$_2$), methanogenesis is thermodynamically compatible with runaway (Appendix~\ref{sec:microbe-thermo}). Furthermore, biology may produce a gas even if it is not metabolically profitable to do so, if that production meets other biological needs. For example, terrestrial biology produces abundant isoprene at considerable energetic cost, because isoprene accomplishes important secondary functions (e.g., stress mitigation; \citealt{Zhan2020}). Finally, modelling of the rise of O$_2$ on Earth suggests that O$_2$, too, underwent runaway, i.e. rapidly and nonlinearly increasing in concentration as a function of surface flux (\citealt{Gregory2021}; Appendix~\ref{sec:runaway_eg}). We therefore argue that the rise of O$_2$ on Earth represents an actualized example of photochemical runaway, affirming its physicality.

Efficient surface deposition (e.g., via biogeochemistry) may prevent biogenic gases from becoming detectable even if they are otherwise able to enter photochemical runaway, but the example of terrestrial O$_2$ suggests that even this barrier can be overcome. Uptake by the surface can limit the concentrations of atmospheric gases. For example, if its surface deposition is efficient, NH$_3$ is restricted to undetectably low concentrations \citep{Huang2020}. Surface deposition can be inefficient; for example, the surface deposition of CO is inefficient due to its insolubility, and CO may enter runaway on M dwarf ocean worlds even if it is deposited at the maximum possible velocity \citep{Kharecha2005, Schwieterman2019}. However, more importantly and more generally, high biogenic gas production can \emph{make} surface deposition inefficient by saturating the surface. This extreme scenario occurred with O$_2$ on Earth. O$_2$ dry deposition into the ocean can theoretically be as high as $10^{-4}~\text{cm}~\text{s}^{-1}$ \citep{Kharecha2005}, which on its own would limit O$_2$ to $20$ ppmv assuming a modern-Earth atmospheric pressure and net O$_2$ production flux. Such oxygen concentrations are extremely low compared to modern oxygen levels of 21\%, and indeed it has been suggested that O$_2$ levels were low for some time after the advent of oxygenic photosynthesis. However, over time, O$_2$ saturated its surface sinks, rendering deposition inefficient and permitting further O$_2$ accumulation \citep{Lyons2014, Daines2017}. We can gain a sense of just how inefficient O$_2$ surface deposition has become by calculating the effective "deposition velocity" required to balance the modern net O$_2$ surface flux in modern Earth's 21\% O$_2$ atmosphere: $v_{dep,eff}=\phi_{O_{2}}/n_{0{2}}=(4\times10^{10}~\text{cm}^{-2}~\text{s}^{-1})/(0.21\times2.5\times10^{19} ~\text{cm}^{-3})=10^{-8} ~\text{cm}~\text{s}^{-1}$, 4 orders of magnitude below the theoretical upper limit. Sustained high production of O$_2$ has saturated O$_2$'s surface biogeochemical sink, rendering it inefficient and permitting O$_2$ accumulation. This is the scenario invoked in for NH$_3$ in the Cold Haber World simulated here. We cannot expect this to happen with every biogenic gas: O$_2$'s modest solubility facilitates saturation, and even so its complete domination of the surface-atmosphere system is limited to the last 0.5-1 Ga \citep{Lyons2014}. However, the possibility that other gases can also saturate their surface sinks and hence suppress their surface deposition rates is not implausible.

Photochemical runaway requires high but biochemically plausible production fluxes, comparable to net O$_2$ production on Earth ($\phi_{O_{2}}=4\times10^{10}~\text{cm}^{-2}~\text{s}^{-1}$  net, \citealt{Zahnle2006}). The magnitude of O$_2$ production is much higher \citep{Gebauer2017}, but most of this O$_2$ is consumed either directly or indirectly by biology; the value we utilize corresponds to production net of these biogenic sinks, as measured by reductant burial and oxidative weathering \citep{Jacob1999, CatlingKasting2017}. It is challenging to predict from first principles which gases are likely to be produced at such high fluxes in exo-biospheres. Even on modern Earth, biogenic gas production is contingent on evolutionary history, and hard to predict from first principles. For example, isoprene is a major product of terrestrial biology, with production rates rivaling those of simple gases associated with primary metabolism (e.g., CH$_4$), despite the fact that isoprene synthesis expends energy and fixed carbon. Rather, isoprene appears to fulfill secondary roles in terrestrial biology, such as stress mitigation. Despite isoprene's high energetic  cost of synthesis, terrestrial production of isoprene is nevertheless high enough to be near the photochemical runaway threshold on planets orbiting M dwarf stars \citep{Zhan2020}. The example of isoprene illustrates the challenge of trying to predict a gas's production from first principles; we argue it is more observationally relevant to be aware of the possibility of diverse biogenic absorbers in exo-atmospheres.

In summary, biogenic gases produced at high but not unrealistic production fluxes can experience nonlinear increases in concentration via the phenomenon of photochemical runaway. Through this phenomenon, even photochemically reactive biosignature gases may accumulate to detectable concentrations, dramatically expanding the palette of potential molecular indicators of life's presence. Photochemical runaway occurred for O$_2$ on Earth; it may occur for other gases on other worlds. Runaways are especially likely on planets orbiting cool stars with low UV output like M dwarfs or white dwarfs. Excitingly, photochemical runaway implies that a world need not develop oxygenic photosynthesis or even terrestrial-type biology in order to have a biosphere that can be remotely detected. Exoplanets to date have proved to be physically more diverse than the worlds of our own solar system; their biologies and their concomitant biosignatures may prove to be similarly diverse \citep{Seager2015}. 

\begin{acknowledgments}
This work was supported in part by a grant from the Simons Foundation (SCOL grant 495062 to S.R.). S.R. gratefully acknowledges the support of Northwestern University's Center for Interdisciplinary Exploration and Research in Astrophysics (CIERA) via the CIERA Fellowship. D.K. was supported by grant 80NSSC20K0269XX from NASA’s XRP program. S.S., Z.Z., W.B., and J.P. received support from the Heising-Simons Foundation (2018-1104). S.S., Z.Z., J.P., and J.H. received supported from NASA (80NSSC19K0471). Z.L. received support from an MIT Presidential Fellowship. This research has made use of NASA’s Astrophysics Data System. We thank B. Gregory and R. Hu for sharing preprints, and we thank B. Gregory, R. Hu, B. Hayworth, C. Harman, S. Rugheimer, R. O. Parke Loyd, and C. Sousa-Silva for discussions. We thank an anonymous referee for detailed feedback that materially improved this paper.

The photochemical model outputs that underlie Figures~\ref{fig:solar_gj876_nh3_runaway} and \ref{fig:solar_gj876_runaway} are available on GitHub at \url{https://github.com/sukritranjan/ranjan_runaway/}. 
The line-by-line climate model used to compute the impact of NH$_3$ and CH$_4$ is available on GitHub: \url{https://github.com/ddbkoll/PyRADS-shortwave}. The radiative transfer model used to simulate transmission spectra is available on GitHub: \url{https://github.com/zhuchangzhan/SEAS}. The scripts used to implement the thermodynamic calculations reported in Appendix~\ref{sec:microbe-thermo} are available on GitHub at \url{https://github.com/sukritranjan/ranjan_runaway/}. The photochemical code used in this paper \citep{Hu2012, Hu2013} is not publicly available, but its input files are available by request. 

\end{acknowledgments}

%

\vspace{5mm}





\appendix
\section{Detailed Model Inputs\label{sec:ap_modeldeets}}
In this section, we present more details regarding the model inputs used in our study.

\subsection{Stellar Spectra}
\begin{figure}[h]
\centering
\includegraphics[width=\linewidth]{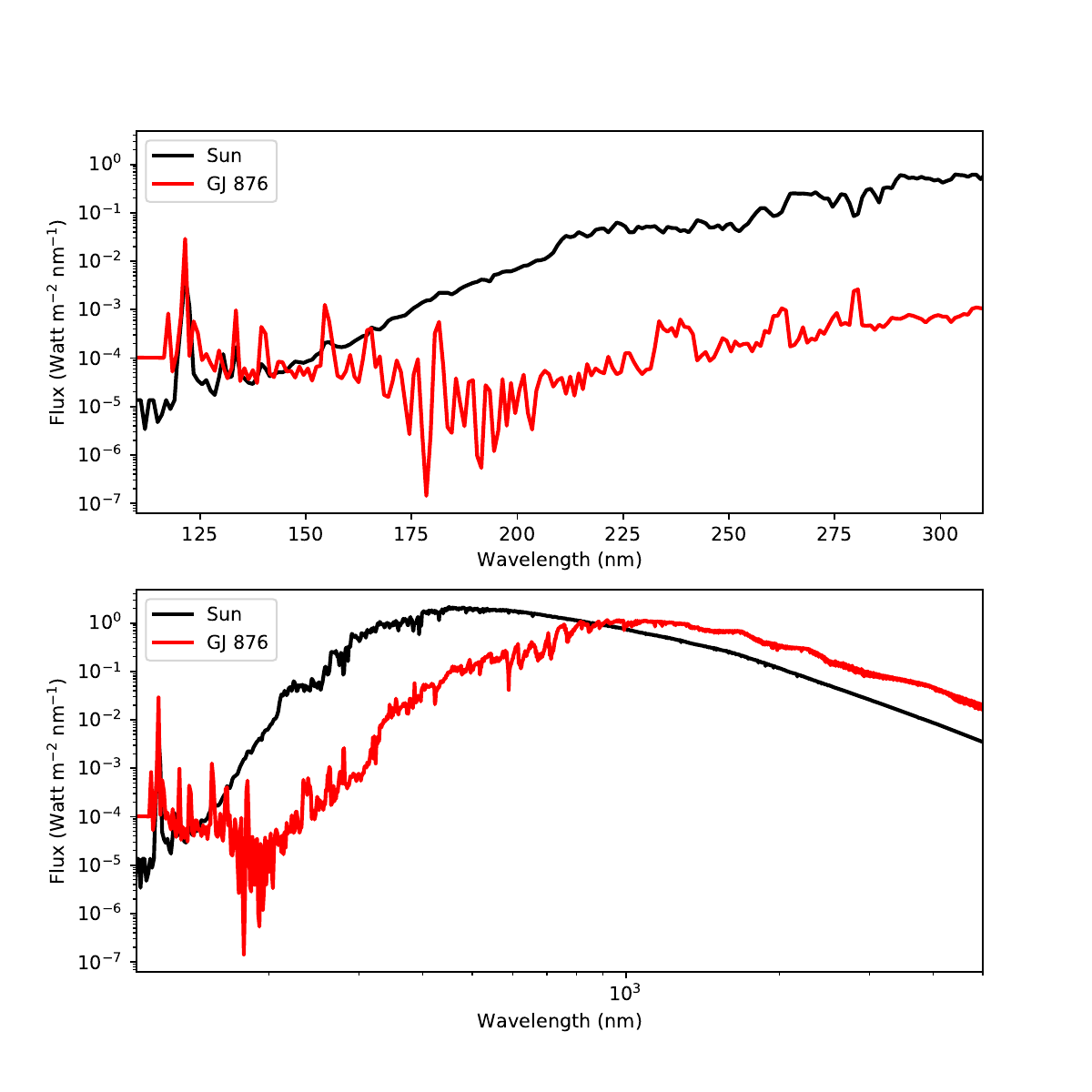}
\caption{Sun and GJ876 spectral irradiation taken as inputs into our photochemical code, for Sun \citep{Hu2012} and GJ 876 irradiation cases \citep{France2016, Youngblood2016, Loyd2016}. The solar spectrum is shown at 1 AU; GJ 876 has been scaled to an equivalent bolometric flux (the same stellar constant). The top plot shows UV irradiation, which is most relevant to photochemistry; the bottom plot shows wider spectral coverage. Small, cool M dwarf stars like GJ 876 emit much less UV radiation compared to Sun-like stars.}
\label{fig:input_stellar_spectra}
\end{figure}

Figure~\ref{fig:input_stellar_spectra} shows the stellar spectra used as inputs in our photochemical model. They are formulated as described in Section~\ref{sec:methods_stellarirrad}, and shown scaled to the modern solar constant (1 AU equivalent).

\subsection{GJ 876 Stellar Parameters\label{sec:ap_stellarparams}}
 We draw stellar parameters for GJ876 (Table~\ref{tab:obs_sim_param}) from the TESS Input Catalog \citep{Stassun2019} and from \citet{Correia2010}. These parameters are used in the simulated JWST observation of the Cold Haber World orbiting GJ 876 (Section~\ref{sec:seas}). To match our assumptions of a planetary instellation equivalent to that received at 1.6 AU orbiting the sun, we assign a planetary semimajor axis of $a=\sqrt{\frac{L_\star}{L_\odot}}\times1.6~\text{AU}=0.18\text{AU}$, in turn implying a period of $P=2\pi\sqrt{\frac{a^{3}}{GM_\star}}=48$ days, an orbital velocity of $v=\sqrt{\frac{GM_\star}{a}}=4.1\times10^4$ m/s, and a transit duration of $T_{transit}=\frac{2R_\star-2R_{planet}}{v} = 3.2$ hours. Here, we have assumed a circular orbit with an impact parameter of zero (negligible orbital inclination). Assuming an impact parameter of zero maximizes the time in transit, meaning that our calculations represent a best-case scenario for number of transits required to detect the trace gas species.
 
 \begin{deluxetable}{lr}
\tablecaption{Stellar parameters used in simulated observation\label{tab:obs_sim_param}}
\tablewidth{0pt}
\tablehead{
\colhead{Parameter} & \colhead{Value} 
 }
\startdata
 $M_\star$, Mass ($M_{sun}$)& 0.34\\
 $R_\star$, Radius ($R_{sun}$) & 0.35 \\
 $L_\star$, Luminosity ($L_{sun}$) & 0.0128 \\
 Effective Temperature (K)& 3271\\
Surface Gravity ($\log_{10}(g)$, cgs) & 4.87\\
$J$-band Magnitude & 5.934 \\
Metallicity (dex) & 0.05 \\ 
\enddata
\tablecomments{Stellar parameters correspond to GJ 876, and are except for metallicity, are drawn from the TESS Input Catalog (TIC) (version 8.1, accessed 08/11/2021, \url{https://exofop.ipac.caltech.edu/tess/target.php?id=188580272}; \citealt{Stassun2019}). Metallicity is drawn from \citet{Correia2010}.} 
\end{deluxetable}

\subsection{Chemical Boundary Conditions}
Table~\ref{tab:species_bcs} presents the detailed chemical boundary conditions used in our baseline simulations of NH$_3$ runaway in an H$_2$-N$_2$ atmosphere. For all species, we assign either a fixed surface mixing ratio or a surface flux. The only species assigned surface mixing ratios are the major atmospheric constituents CO$_2$, N$_2$, and H$_2$, depending on the planet scenario, and H$_2$O, set based on surface temperature and an assumed relative humidity following \citet{Hu2012}. For species with a surface flux, the species is assumed to be injected at the bottommost layer in the atmosphere. 

\begin{longrotatetable}
\begin{deluxetable}{cccccccp{4cm}}
\tablecaption{General species boundary conditions\label{tab:species_bcs}}
\tablewidth{0pt}
\tablehead{
\colhead{Species\tablenotemark{a}} & \colhead{Type} & \colhead{$\phi_{esc}$\tablenotemark{b}} & \colhead{$r_{surf}$\tablenotemark{c}} & \colhead{$\phi_{surf}$\tablenotemark{d}}  & \colhead{$v_{dep, low}$\tablenotemark{e}} & \colhead{$v_{dep,high}$\tablenotemark{f}} & \colhead{$v_{dep}$ Choice Notes}\\
  & & \colhead{(cm$^{-2}$ s$^{-1}$)} & & \colhead{(cm$^{-2}$ s$^{-1}$)} & \colhead{(cm s$^{-1}$)} & \colhead{(cm s$^{-1}$)} &
 }
\startdata
N$_2$ & C & 0 & 0.1 & -- & -- & -- &\\
CO$_2$ & X & 0 & -- & $3.0\times10^{11}$ & $1\times10^{-4}$ & $1\times10^{-4}$ & \cite{Hu2012}\\
H$_2$ & X & Diff.-limit. & 0.9 & -- & -- & -- & \\
NO & X & 0 & -- & 0 & $3\times10^{-4}$& 0.02& $v_{dep, low}$: \cite{Harman2015}. $v_{dep, high}$: \cite{Hu2012, SeinfeldPandis} (Earth continent) \\
CO & X & 0 & -- & 0 & $1\times10^{-8}$ & $0.03$&$v_{dep, low}$: \cite{Kharecha2005, Hu2012} (abiotic ocean). $v_{dep, high}$: 0.03 \cite{Hu2012, SeinfeldPandis} (Earth continent)\\ 
O$_2$ & X & 0 & -- & 0 & 0 & $1\times10^{-4}$&$v_{dep, low}$: \cite{Hu2012} (no surface sink). $v_{dep, high}$:\cite{Domagal-Goldman2011, Harman2015} (oceanic piston velocity-limited)\\
CH$_4$ & X & 0 & -- & $3\times10^{8}$ & 0 & $1\times10^{-4}$ & $v_{dep, low}$: \cite{Hu2012, Harman2015} (no surface sink). $v_{dep, high}$: \cite{Hu2012} (Earthlike) \\ 
\textbf{NH$_3$} & \textbf{X} & \textbf{0} & \textbf{--} & \textbf{variable} & \textbf{$1\times10^{-7}$} & \textbf{$1\times10^{-7}$}& Prescribed (c.f. Section~\ref{sec:planet_scenarios}) \\
\cline{1-8}\\
H & X & Diff.-limit. & -- & 0 & 1 & 1 &\cite{Hu2012, Harman2015} \\
O & X & 0 & -- & 0 & 1 & 1 &\cite{Hu2012, Harman2015}\\
O(1D) & X & 0 & -- & 0 & 0 & 1 &$v_{dep, low}$: \cite{Hu2012, Harman2015}\tablenotemark{g} (abiotic). $v_{dep, high}$: \cite{Hu2012} (Earthlike)\\
O$_3$ & X & 0 & -- & 0 & 0.4 & 0.4&\cite{Hu2012}\\
OH & X & 0 & -- & 0 & 1 & 1 &\cite{Hu2012}\\
HO$_2$ & X & 0 & -- & 0 & 1 & 1 &\cite{Hu2012}\\
H$_2$O & X & 0 & 0.01 & -- & -- & -- &\cite{Hu2012}\\
H$_2$O$_2$ & X & 0 & -- & 0 & 0.5 & 0.5 &\cite{Hu2012}\\
CH$_2$O & X & 0 & -- & 0 &0.1 & 0.2&$v_{dep, low}$:\cite{Hu2012}. $v_{dep, high}$:\cite{Domagal-Goldman2011}\\
CHO & X & 0 & -- & 0 &0.1 & 1 &$v_{dep, low}$: \cite{Hu2012}.$v_{dep, high}$:\cite{Domagal-Goldman2011}\\
C & X & 0 & -- & 0 & 0 & 0 &\cite{Hu2012}\\
CH & X & 0 & -- & 0 & 0 & 0 &\cite{Hu2012}\\
CH$_2$ & X & 0 & -- & 0 & 0 & 0 &\cite{Hu2012}\\
CH$_2^1$ & X & 0 & -- & 0 & 0 & 0&\cite{Hu2012}\\
CH$_3$ & X & 0 & -- & 0 & 0 & 1 &$v_{dep, low}$:\cite{Hu2012}. $v_{dep, high}$:\cite{Domagal-Goldman2011}\\
CH$_3$O & X & 0 & -- & 0 & 0.1 & 0.1 &\cite{Hu2012}\\
CH$_4$O & X & 0 & -- & 0 & 0.1 & 0.1 &\cite{Hu2012}\\
CHO$_2$ & X & 0 & -- & 0 & 0.1 & 0.1 &\cite{Hu2012}\\
CH$_2$O$_2$ & X & 0 & -- & 0 & 0.1 & 0.5 &$v_{dep, low}$:\cite{Hu2012} (abiotic). $v_{dep, high}$: \cite{Hu2012} (Earthlike)\\
CH$_3$O$_2$ & X & 0 & -- & 0 & 0.1 & 1 &$v_{dep, low}$:\cite{Hu2012} (abiotic). $v_{dep, high}$: \cite{Hu2012} (Earthlike)\\
CH$_4$O$_2$ & X & 0 & -- & 0 & 0.1 & 0.1 &\cite{Hu2012}\\
C$_2$ & X & 0 & -- & 0 & 0 & 0 &\cite{Hu2012}\\
C$_2$H & X & 0 & -- & 0 & 0 & 0 &\cite{Hu2012}\\
C$_2$H$_2$ & X & 0 & -- & 0 & 0 & 0 &\cite{Hu2012}\\
C$_2$H$_3$ & X & 0 & -- & 0 & 0 & 0 &\cite{Hu2012}\\
C$_2$H$_4$ & X & 0 & -- & 0 & 0 & 0 &\cite{Hu2012}\\
C$_2$H$_5$ & X & 0 & -- & 0 & 0 & 0 &\cite{Hu2012}\\
C$_2$H$_6$ & X & 0 & -- & 0 & $1\times10^{-5}$ & $1\times10^{-5}$ &\cite{Hu2012}\\
C$_2$HO & X & 0 & -- & 0 & 0.1 & 0.1&\cite{Hu2012}\\
C$_2$H$_2$O & X & 0 & -- & 0 & 0.1 & 0.1&\cite{Hu2012}\\
C$_2$H$_3$O & X & 0 & -- & 0 & 0.1 & 0.1 &\cite{Hu2012}\\
C$_2$H$_4$O & X & 0 & -- & 0 & 0.1 & 0.1 &\cite{Hu2012}\\
C$_2$H$_5$O & X & 0 & -- & 0 & 0.1 & 0.1 &\cite{Hu2012}\\
S & X & 0 & -- & 0 & 0 & 1 &$v_{dep, low}$:\cite{Hu2012}. $v_{dep, high}$:\cite{Domagal-Goldman2011}\\
S$_2$ & X & 0 & -- & 0 & 0 & 0 &\cite{Hu2012}\\
S$_3$ & X & 0 & -- & 0 & 0 & 0 &\cite{Hu2012}\\
S$_4$ & X & 0 & -- & 0 & 0 & 0 &\cite{Hu2012}\\
SO & X & 0 & -- & 0 & 0 & $3\times10^{-4}$&\cite{Hu2012, Domagal-Goldman2011}\\
SO$_2$ & X & 0 & -- & $3\times10^{9}$ & 1 & 1 &\cite{Hu2012}\\
SO$_2^1$ & X & 0 & -- & 0 & 0 & 1 &$v_{dep, low}$:\cite{Hu2012} (abiotic). $v_{dep, high}$: \cite{Hu2012} (Earthlike)\\
SO$_2^3$ & X & 0 & -- & 0 & 0 & 1 &$v_{dep, low}$:\cite{Hu2012} (abiotic). $v_{dep, high}$: \cite{Hu2012} (Earthlike)\\
SO$_3$ & X & 0 & -- & 0 & 1 & 1 &\cite{Hu2012}\\
H$_2$S & X & 0 & -- & $3\times10^{8}$ & 0.015 & 0.015&\cite{Hu2012}\\
HS & X & 0 & -- & 0 & 0 & 1 &$v_{dep, low}$:\cite{Hu2012}. $v_{dep, high}$:\cite{Domagal-Goldman2011}\\ 
HSO & X & 0 & -- & 0 & 0 & 1 &$v_{dep, low}$:\cite{Hu2012}. $v_{dep, high}$:\cite{Domagal-Goldman2011}\\
HSO$_2$ & X & 0 & -- & 0 & 0 & 0.1 &$v_{dep, low}$:\cite{Hu2012} (abiotic). $v_{dep, high}$: \cite{Hu2012} (Earthlike)\\
HSO$_3$ & X & 0 & -- & 0 & 0.1 & 0.1 &\cite{Hu2012}\\
H$_2$SO$_4$ & X & 0 & -- & 0 & 1 & 1 &\cite{Hu2012}\\
H$_2$SO$_4$(A) & A & 0 & -- & 0 & 0.2 & 0.2 &\cite{Hu2012}\\
S$_8$ & X & 0 & -- & 0 & 0 & 0.2 &$v_{dep, low}$:\cite{Hu2012} (abiotic). $v_{dep, high}$: \cite{Hu2012} (Earthlike)\\
S$_8$(A) & A & 0 & -- & 0 & 0.2 & 0.2 &\cite{Hu2012}\\
OCS & X & 0 & -- & 0 & 0.01 & 0.01 &\cite{Hu2012} (Earthlike)\\
CS & X & 0 & -- & 0 &  0 & $1\times10^{-4}$ &$v_{dep, low}$:\cite{Hu2012}. $v_{dep, high}$:\cite{Domagal-Goldman2011}\\
CH$_3$S & X & 0 & -- & 0 & 0.01 & 0.01 &\cite{Hu2012}\\ 
CH$_4$S & X & 0 & -- & 0 & 0.01 & 0.01 &\cite{Hu2012}\\
N & X & 0 & -- & 0 & 0 & 0 &\cite{Hu2012}\\
NH$_2$ & X & 0 & -- & 0 & 0 & 1 &$v_{dep, low}$:\cite{Hu2012}. $v_{dep, high}$:\cite{Domagal-Goldman2011}\\
NH & X & 0 & -- & 0 & 0 & 0 &\cite{Hu2012}\\
N$_2$O & X & 0 & -- & 0 & 0 & 0 &\cite{Hu2012, Hu2019}\\
NO$_2$ & X & 0 & -- & 0 &  $3\times10^{-3}$& 0.02& $v_{dep, low}$:\cite{Domagal-Goldman2011}. $v_{dep, high}$:\cite{Hu2012}\\
NO$_3$ & X & 0 & -- & 0 & 1 & 1&\cite{Hu2012}\\
N$_2$O$_5$ & X & 0 & -- & 0 & 1 & 4&$v_{dep, low}$:\cite{Hu2012, Hu2019} (abiotic). $v_{dep, high}$: \cite{Hu2012} (Earthlike)\\
HNO & X & 0 & -- & 0 & 0 & 1 &$v_{dep, low}$:\cite{Hu2012}. $v_{dep, high}$:\cite{Domagal-Goldman2011}\\
HNO$_2$ & X & 0 & -- & 0 & 0.5 & 0.5 &\cite{Hu2012}\\
HNO$_3$ & X & 0 & -- & 0 & 1 & 4&$v_{dep, low}$:\cite{Hu2012, Hu2019} (abiotic). $v_{dep, high}$: \cite{Hu2012} (Earthlike)\\
HNO$_4$ & X & 0 & -- & 0 & 1 & 4&$v_{dep, low}$:\cite{Hu2012, Hu2019} (abiotic). $v_{dep, high}$: \cite{Hu2012} (Earthlike)\\
HCN & X & 0 & -- & 0 & $7\times10^{-3}$ &  0.01& $v_{dep, low}$:\cite{Tian2011}. $v_{dep, high}$:\cite{Hu2012}\\
CN & X & 0 & -- & 0 & 0.01 & 1&$v_{dep, low}$:\cite{Hu2012}. $v_{dep, high}$:\cite{Tian2011}\\
CNO & X & 0 & -- & 0 & 0 & 1&$v_{dep, low}$:\cite{Hu2012}. $v_{dep, high}$:\cite{Tian2011}\\
HCNO & X & 0 & -- & 0 & 0 & 1&$v_{dep, low}$:\cite{Hu2012}. $v_{dep, high}$:\cite{Tian2011}\\
CH$_3$NO$_2$ & X & 0 & -- & 0 & 0.01 & 0.01&\cite{Hu2012}\\
CH$_3$NO$_3$ & X & 0 & -- & 0 & 0.01 & 0.01&\cite{Hu2012}\\
CH$_5$N & X & 0 & -- & 0 & 0 & 1.0 &$v_{dep, low}$:\cite{Hu2012}. $v_{dep, high}$: assigned as 1 based on solubility (c.f. \cite{Hu2012})\\
C$_2$H$_2$N & X & 0 & -- & 0 & 0 & 0.02 &$v_{dep, low}$:\cite{Hu2012}. $v_{dep, high}$: assigned as 0.02 as a reactive radical following \cite{Hu2012}\\
N$_2$H$_2$ & X & 0 & -- & 0 & 0 & 1 &$v_{dep, low}$:\cite{Hu2012}. $v_{dep, high}$: assigned as 1 following N$_2$H$_4$ (below)\\
N$_2$H$_3$ & X & 0 & -- & 0 & 0 & 1 &\cite{Hu2012}. $v_{dep, high}$:\cite{Domagal-Goldman2011}\\
N$_2$H$_4$ & X & 0 & -- & 0 & 0 & 1 &$v_{dep, low}$:\cite{Hu2012}. $v_{dep, high}$: assigned as 1 based on solubility (c.f. \cite{Hu2012})\\
\enddata
\tablenotetext{a}{``X": the full continuity-diffusion equation is solved for the species; ``A": aerosol, settles out of the atmosphere; ``C": chemically inert. Our code also has the capability to designate species as type``F", i.e.  treated as being in photochemical equilibrium \citep{Hu2012}. However, we do not consider any type ``F" species in this work to better enforce mass balance \citep{James2018}.}
\tablenotetext{b}{Escape flux from the TOA; a negative number corresponds to inflow.}
\tablenotetext{c}{Surface mixing ratio relative to dry air}
\tablenotetext{d}{Surface emission flux}
\tablenotetext{e}{Dry deposition velocity, standard case}
\tablenotetext{f}{Dry deposition velocity, high-deposition sensitivity test}
\tablenotetext{g}{Species treated as being in photochemical equilibrium have an implicit 0 deposition velocity}
\tablecomments{For the bottom boundary condition, either the surface mixing ratio or the surface flux and deposition velocity are specified.}
\end{deluxetable}
\end{longrotatetable}

\section{NH$_3$ Photochemical Runaway in Detail \label{sec:runaway_nh3} }
We discuss here the runaway of NH$_3$ in the Cold Haber World scenario in more detail, including sensitivity tests. NH$_3$ runaway has not been previously reported in the literature. The solubility of NH$_3$ means that in most terrestrial planet contexts, wet deposition (rainout) will regulate pNH$_3$ and photochemistry is largely irrelevant \citep{Huang2020}. However, if the surface can be saturated in NH$_3$, e.g. due to a biosphere that is a net source of NH$_3$ as invoked in the Cold Haber World scenario we simulate, then photochemistry controls pNH$_3$ \citep{Kasting1982, Seager2013, Seager2013b}. In this case, we find the main NH$_3$ loss mechanism to be direct photolysis, as found by \citet{Kuhn1979} and  \citet{Kasting1982} for N$_2$-dominated atmospheres (early Earth). We found photolysis to dominate other photochemical loss processes by orders of magnitude. Due to the generally lower UV output of M dwarfs at the NH$_3$-photolyzing $\leq230$ nm wavelengths, NH$_3$ enters runaway and becomes detectable for M dwarfs at fluxes for which it is still firmly photochemically limited to undetectable concentrations for solar-type stars. 

\begin{figure}[h]
\centering
\includegraphics[width=\linewidth]{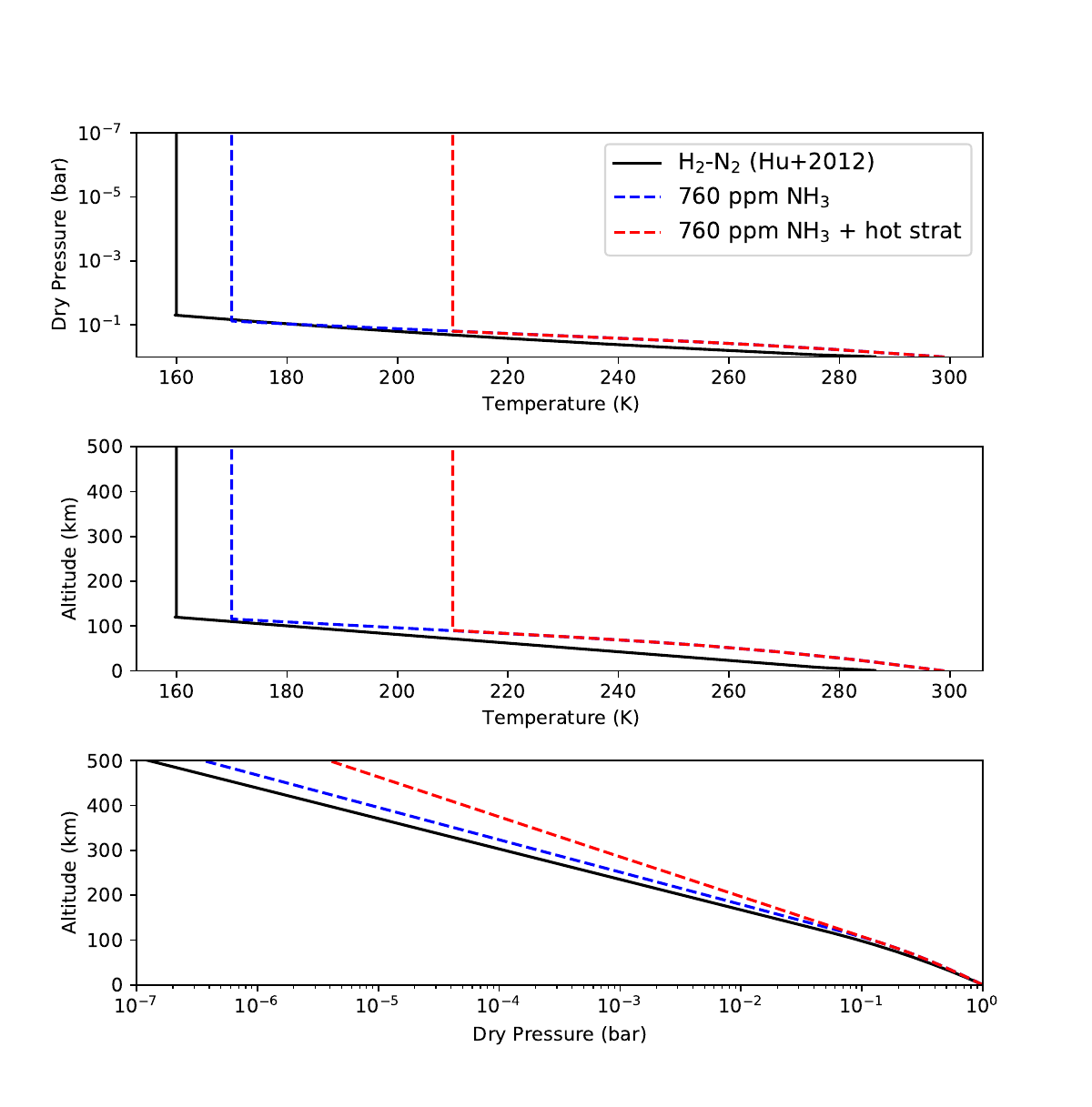}
\caption{Temperature-pressure profiles for Cold Haber World simulations. The black line refers to the standard profiles, drawn from \cite{Hu2012}. The blue dashed line refers to a planet with $T_{surf}=299$K and $T_{strat}=170$K, to account for the greenhouse effect of NH$_3$. The red dashed line refers to a planet with $T_{surf}=299$K and $T_{strat}=210$K, to account for the potential shortwave heating driven by abundant NH$_3$ and its photochemical products}
\label{fig:tp_sens_test}
\end{figure}

We considered the impact of NH$_3$ greenhouse warming on NH$_3$ runaway. For the H$_2$-dominated Cold Haber World planetary scenario we study here, we found the greenhouse warming due to NH$_3$ to be relatively modest, because NH$_3$ absorption near the peak of the planet's outgoing longwave radiation (OLR) occurs at similar wavelengths as the strong H$_2$-H$_2$ collision-induced absorption (CIA) in our H$_2$-dominated scenario. Specifically, we found that adding 55 ppmv NH$_3$ (corresponding approximately to the 70 ppmv predicted in our base case calculations for $\phi_{NH_{3}}=2\times10^{10}$ cm$^{-2}$ s$^{-1}$) increased the surface temperature $T_{surf}$ by only 11K, from 288K to 299K. At this abundance, NH$_3$ is already detectable with an average significance of $5\sigma$ in 2 transits from JWST. We constructed a new temperature-pressure profile evolving as a wet adiabat from $T_{surf}=299$ K to an isothermal stratosphere of $T_{strat}=170$ K (Figure~\ref{fig:tp_sens_test}), and reran our simulations, adjusting our surface H$_2$O molar mixing ratio boundary condition from 0.01 to 0.02 to account for higher H$_2$O saturation pressure. We found a negligible impact of higher surface temperature on NH$_3$ accumulation. Note that the main effect of higher $T_{surf}$ is to increase the water vapor content, and hence H and OH production rates, in the lower atmosphere. As the main photochemical sink of NH$_3$ is direct photolysis at higher altitudes, NH$_3$ accumulation is not strongly affected by changes in lower-atmosphere H$_2$O abundance.


We considered the possibility of higher stratospheric temperatures, and their potential effect on NH$_3$ accumulation. NH$_3$ and its photochemical product N$_2$H$_4$ are strong UV absorbers, and we considered the possibility that their absorption of UV might warm the stratosphere, as does O$_3$ on Earth. To conduct a sensitivity test for this possibility, we constructed a temperature-pressure profile evolving as a wet adiabat from $T_{surf}=299$ K to an isothermal stratosphere of $T_{strat}=210$ K (Figure~\ref{fig:tp_sens_test}), and reran our simulations, adjusting our surface H$_2$O mixing ratio boundary condition from 0.01 to 0.02 to account for higher H$_2$O saturation pressure. We chose $T_{strat}=210$ K based on modern Earth's empirical tropopause temperature, which is significantly warmer than the $T_{strat}$ implied by our climate calculations and thus should provide a conservative upper bound.
We found that warmer stratospheres lead to much higher NH$_3$ abundances for the same $\phi_{NH_{3}}$, i.e. warmer stratospheres promoted the accumulation of NH$_3$. We attribute this to the stabilization of NH$_3$ due to higher production of H driven by higher stratospheric H$_2$O abundances. In more detail: \citet{Hu2012} report the main source of H in the atmospheres of habitable worlds with H$_2$-N$_2$ atmospheres to be H$_2$O photolysis, via
\begin{eqnarray}
H_2O + h\nu \rightarrow H + OH\\
OH + H_2 \rightarrow H_2O + H
\end{eqnarray}
Therefore, increasing stratospheric H$_2$O increases H production. \citet{Kasting1982} reports that increased H stabilizes NH$_3$ by aiding the recombination $NH_2 + H +M \rightarrow NH_3$; we observe the same. \citet{Kasting1982} also indicated that the removal of N$_2$H$_3$ stabilizes NH$_3$, since N$_2$H$_3$ is the key intermediary en route from NH$_3$ to N$_2$. Higher H suppresses N$_2$H$_3$ via $N_2H_3 + H \rightarrow 2NH_2$, stabilizing NH$_3$. H also suppresses N$_2$H$_2$ and N$_2$H$_4$, which are other intermediaries in the oxidation of NH$_3$ to N$_2$. Therefore, our finding that hotter stratospheres stabilize NH$_3$ is consistent with past work. Indeed, we observed higher NH$_3$ and lower NH$_2$, N$_2$H$_2$, NH$_2$H$_3$, and N$_2$H$_4$ in our "hot stratosphere" simulations compared to our standard simulations, as expected by this theory.

At high NH$_3$ concentrations, our model predicts elevated concentrations of N$_2$H$_4$ (hydrazine), which is condensable at habitable temperatures. This raises the possibility of hydrazine haze in high-NH$_3$ atmospheres. We compared the hydrazine concentrations predicted by our model to the saturation pressure of hydrazine in equilibrium with the solid phase, using the expression of \citet{Atreya1977}. We found hydrazine to be condensable in the upper atmosphere if the stratosphere is cold ($T_{strat}=160$ K), but not if the stratosphere is warm ($T_{strat}\geq170$ K). This is both because the condensation pressure of N$_2$H$_4$ is higher at higher temperatures, and because warmer stratospheres inhibit N$_2$H$_4$ accumulation, as above. Since the high NH$_3$ abundances that may produce condensable N$_2$H$_4$ should also produce warmer stratospheric temperatures (see \citealt{Rugheimer2018} for CH$_4$), it is unclear whether haze should in fact form in high-NH$_3$ atmospheres; coupled climate-photochemistry calculations are required to resolve this ambiguity. If N$_2$H$_4$ haze does form, then it might inhibit detection of spectral features of NH$_3$ and other molecules in transmission spectra, as has been observed in large gaseous exoplanet atmospheres \citep{Deming2017}. On the other hand, N$_2$H$_4$ ice has spectral features that may enable its unique identification, in which case N$_2$H$_4$ haze could be used as a probe of NH$_3$, comparable to the way in which organic haze has been suggested as a probe of organic gas production \citep{Zheng2008, Arney2018}. 

\begin{figure}[h]
\centering
\includegraphics[width=\linewidth]{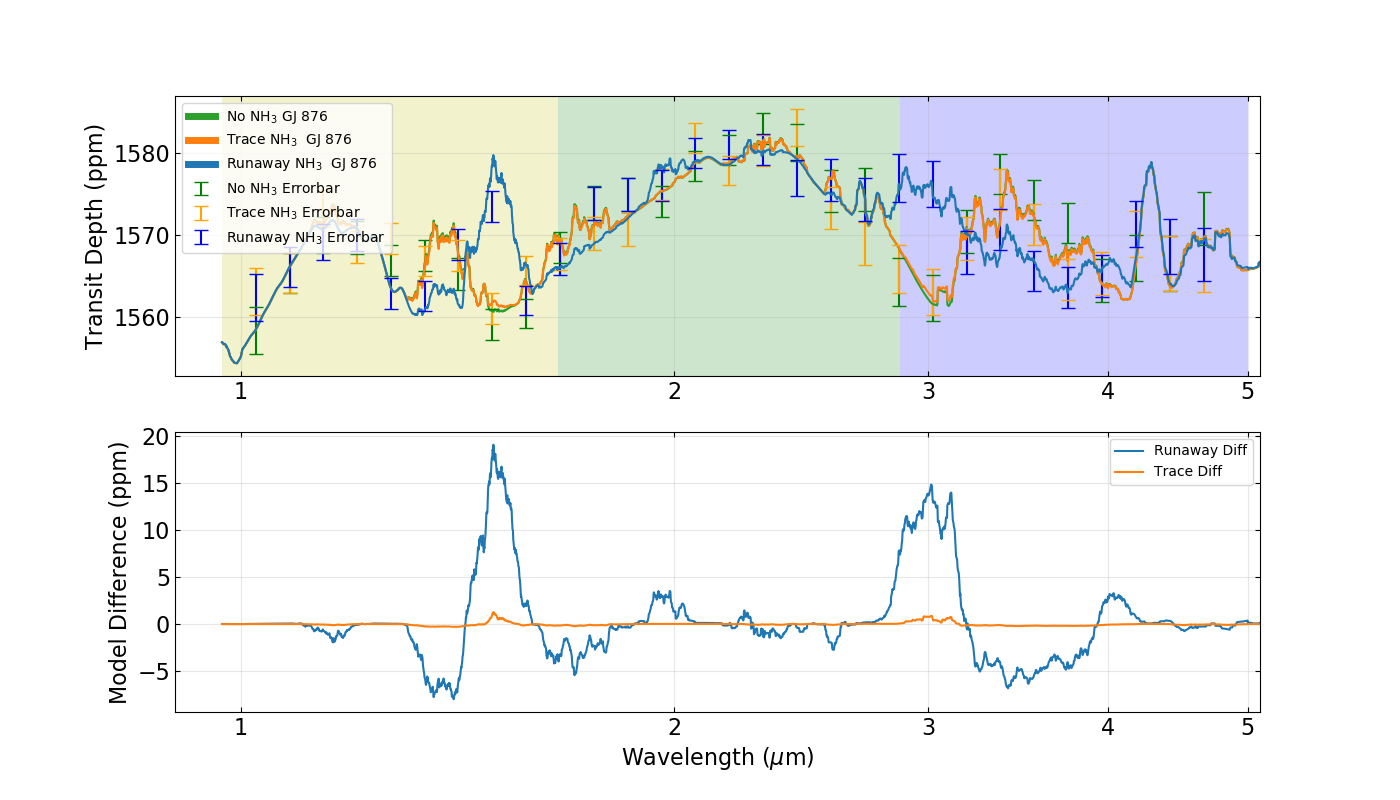}
\caption{Sample simulated detection of NH$_3$ with JWST. Upper panel: simulated observation of the exoplanet transmission spectrum ($R=10$, two transits) for a rocky planet with an H$_2$-dominated atmosphere transiting an M dwarf star for NH$_3$ surface fluxes of 2$\times$ 10$^{10}~\text{cm}^{-2}~\text{s}^{-1}$ (blue),  3 $\times$ 10$^{9}~\text{cm}^{-2}~\text{s}^{-1}$ (orange), and $0~\text{cm}^{-2}~\text{s}^{-1}$ (green). The y-axis shows transit depth (ppm), and the x-axis
shows wavelength ($\mu$m). The spectra are simulated from 0.3 to 5 $\mu$m, covering the wavelength span of the JWST NIRSpec instrument. The yellow, green, and blue region shows the spectral coverage of NIRSpec 140 M, 235 M, and 395 M, respectively. Lower panel: the difference between the $0~\text{cm}^{-2}~\text{s}^{-1}$ and 3 $\times$ 10$^{9}~\text{cm}^{-2}~\text{s}^{-1}$ atmosphere (orange), and the $0~\text{cm}^{-2}~\text{s}^{-1}$ and 2 $\times$ 10$^{10}~\text{cm}^{-2}~\text{s}^{-1}$ atmosphere (blue). This figure is one of a suite of 100 simulated observations conducted for this study; in this simulated observation, the NH$_3$ $1.45-1.55\mu$m feature is detected at 5.4$\sigma$, whereas the 100-simulation mean is $5.0\sigma$. NH$_3$ can be detected for the atmosphere in NH$_3$ runaway via the $1.45-1.55\mu$m feature.}
\label{fig:snr_nh3}
\end{figure}

We now turn to the effect of photochemical runaway on the detectability of NH$_3$. We compare simulated observations of our 1.5 $R_{Earth}$, 5 $M_{Earth}$ Cold Haber World with $\phi_{NH_{3}}=0$ cm$^{-2}$ s$^{-1}$, $\phi_{NH_{3}}=3\times10^{9}$ cm$^{-2}$ s$^{-1}$ (not in NH$_3$ runaway), and $\phi_{NH_{3}}=2\times10^{10}$ cm$^{-2}$ (NH$_3$ runaway) binned to $R=10$ in Figure~\ref{fig:snr_nh3}. The $\phi_{NH_{3}}=3\times10^{9}$ cm$^{-2}$ s$^{-1}$ simulated observation cannot be differentiated from the $\phi_{NH_{3}}=0$ cm$^{-2}$ s$^{-1}$ observation with JWST; the JWST noise floor exceeds the difference between the models. On the other hand, the $\phi_{NH_{3}}=2\times10^{10}$ cm$^{-2}$ s$^{-1}$ simulated observation can be be differentiated from the $\phi_{NH_{3}}=0$ cm$^{-2}$ s$^{-1}$ simulated observation with two JWST transits. Specifically, in a suite of 100 simulated observations\footnote{The simulation parameters are identical; the simulations differ only in their randomly generated noise.} of two transits of our planetary scenario with JWST, the $1.45-1.55\mu$m NH$_3$ feature can be distinguished with an average significance of $5.0\sigma$. The $2.85-3.15\mu$m NH$_3$ feature is not robustly detected with two transits (average significance $2.24\sigma$), and requires additional transit observations. 

\section{Photochemical Runaway of CO, O$_2$, and CH$_4$ \label{sec:runaway_eg}}
We illustrate the generality of the phenomenon of runaway by calculating the concentration as a function of surface production flux for three gases other than NH$_3$ using our photochemical model (see Tables~\ref{tab:planet_param_eg}, \ref{tab:scenario_species_bcs}). We consider the accumulation of CH$_4$ in a CO$_2$-dominated atmosphere and the accumulation of CO and O$_2$ in N$_2$-dominated atmospheres, for both solar and M dwarf irradiation (represented by GJ 876), to complement our earlier simulation of NH$_3$ accumulation in an H$_2$-dominated atmosphere. We chose these gases because they are well studied in the literature, meaning that their reaction networks are relatively complete, and because runaway of these gases has been studied in previous work (albeit not as a unified concept) \citep{Kasting1990, Pavlov2003, Segura2005, Zahnle2006, Schwieterman2019, Gregory2021}. We made these choices to demonstrate that the phenomenon of runaway is not necessarily attributable to incomplete reaction networks, nor is it unique to our model. We similarly chose background atmospheres of varying redox state to illustrate the generality of the phenomenon. Unless otherwise stated, we used the same model choices as for NH$_3$ runaway, to illustrate the recovery of the runaway phenomenon for diverse gases with a uniform model calculation. 

\subsection{Methods}
To demonstrate the generality of runaway, we model the runaway of CO, O$_2$, and CH$_4$ in N$_2$- or CO$_2$-dominated atmospheres. Such runaways have been reported previously in the literature, though they have not always been identified as such. To carry out these calculations, we use the same photochemical model and reaction network as for the NH$_3$ runaway. We include a lightning-produced NO flux of $1\times10^{8}$ cm$^{-2}$ s$^{-1}$ injected at the surface, and an equal conjugate CO production, for redox balance \citep{Harman2018, Hu2019}. For an Earth-sized planet, these correspond to global mass fluxes of 0.8 Tg NO yr$^{-1}$ and 0.8 Tg CO year$^{-1}$; this is lower than modern Earth, because CO$_2$ is a less efficient oxidant of N$_2$ than O$_2$ \citep{Harman2018}. We adopt vertical temperature, pressure, and eddy diffusion profiles from \citet{Hu2012}, following the logic detailed therein. We follow \citet{Hu2012} in adopting semimajor axes for the N$_2$ and CO$_2$ atmospheres of 1.0 and 1.3  for the solar instellation case, based on crude climate calculations, and we scale our GJ 876 spectra to the same total instellation as in the corresponding Solar case for each atmospheric scenario. 

For chemical boundary conditions, we modify the low deposition velocity boundary conditions given in Table~\ref{tab:species_bcs}, as indicated in Table~\ref{tab:scenario_species_bcs}. As these instances of runaway have been previously reported, we consider them robust, and do not carry out the high deposition velocity sensitivity test executed for NH$_3$. For these CO$_2$- and N$_2$-dominated atmospheric scenarios, we employ the additional solution quantity check of verifying the atmospheric redox balance \citep{Harman2015, James2018}. We are unable to employ this quality check for the H$_2$-dominated atmospheric scenario studied for NH$_3$ runaway in the main paper because a fixed H$_2$ mixing ratio boundary condition is employed in this case, and our code is incapable of calculating atmospheric redox balance when a fixed mixing ratio boundary condition is employed with a species with a non-neutral redox state.

To simulate CO runaway, we chose an atmospheric scenario motivated by early Earth, for which CO runaway has been extensively studied \citep{Kasting1990, Kasting2014, Schwieterman2019, Ranjan2020}. We chose an N$_2$-dominated atmosphere, and adopted a CO$_2$ mixing ratio of 0.01. We selected a low CO$_2$ mixing ratio to ensure that photolytic production of CO from CO$_2$ photolysis did not tip the atmosphere into runaway on its own, whether orbiting a Sun-like star or an M dwarf \citep{Kasting2014, Hu2020}. 

To simulate O$_2$ runaway, we again chose an atmospheric scenario motivated by early Earth, which experienced O$_2$ buildup \citep{Gregory2021}. We chose an N$_2$-dominated atmosphere, and adopted a CO$_2$ mixing ratio of 0.01. We selected a low CO$_2$ mixing ratio to ensure that photolytic production of O$_2$ did not tip the atmosphere into runaway on its own \citep{Hu2020}. 

To simulate CH$_4$ runaway, we chose a CO$_2$-dominated atmospheric scenario, because our model is not equipped to handle the hydrocarbon haze formation that accompanies high CH$_4$/CO$_2$ ratios. We adopt a CO$_2$ mixing ratio of 0.9 and an N$_2$ mixing ratio of 0.1. 

 \begin{deluxetable}{lccc}
\tablecaption{Planetary Parameters Assumed in the Photochemical Calculation of the CO, O$_2$, and CH$_4$ Runaways\label{tab:planet_param_eg}}
\tablewidth{0pt}
\tablehead{
\colhead{Parameter} & \colhead{CO Runaway} & \colhead{O$_2$ Runaway} & \colhead{CH$_4$ Runaway}
 }
\startdata
Bulk Composition & $99$\% N$_2$, 1\% CO$_2$ & $99$\% N$_2$, 1\% CO$_2$ & 90\% CO$_2$, 10\% N2$_2$\\
Surface Water Vapor Mixing Ratio & 0.01 & 0.01 & 0.01\\
Surface Pressure (bar) & 1 & 1 & 1 \\
Surface Temperature (K) & 288 & 288 & 288 \\
Stratospheric Temperature (K) & 200 & 200 & 175 \\
Stellar Irradiation Considered & Sun, GJ 876 & Sun, GJ 876 & Sun, GJ 876\\
Stellar Constant Relative to Modern Earth & 1 & 1 & 0.6\\
Planet Mass  ($M_{\Earth}$)& 1 & 1 & 1\\
Planet Radius ($R_{\Earth}$) & 1 & 1 & 1\\
\enddata
\tablecomments{Bulk planetary properties assumed in modeling the CO, O$_2$, and CH$_4$ runaways. Planet scenarios are based on the N$_2$-dominated and CO$_2$-dominated exoplanet atmospheric benchmark scenarios of \citet{Hu2012}.} 
\end{deluxetable}

\begin{longrotatetable}
\begin{deluxetable}{ccccccp{7cm}}
\tablecaption{This Table updates the boundary conditions presented in Table~\ref{tab:species_bcs} for the other runaway scenarios studied in Appendix~\ref{sec:runaway_eg}.\label{tab:scenario_species_bcs}}
\tablewidth{0pt}
\tablehead{
\colhead{Species\tablenotemark{a}} & \colhead{Type} & \colhead{$\phi_{esc}$\tablenotemark{b}} & \colhead{$r_{surf}$\tablenotemark{c}} & \colhead{$\phi_{surf}$\tablenotemark{d}}  & \colhead{$v_{dep, low}$\tablenotemark{e}} &  \colhead{$v_{dep}$ Choice Notes}\\
  & & \colhead{(cm$^{-2}$ s$^{-1}$)} & & \colhead{(cm$^{-2}$ s$^{-1}$)} & \colhead{(cm s$^{-1}$)}  &
 }
\startdata
\multicolumn{6}{c}{CO Runaway in N$_2$-CO$_2$ Atmosphere}\\
\cline{2-5}\\
N$_2$ & C & 0 & 0.99 & -- & -- &\\
CO$_2$ & X & 0 & 0.01 & -- & -- & \\
H$_2$ & X & Diff.-limit. & -- & $3.\times10^{10}$ & 0 &\cite{Hu2012} \\
NO & X & 0 & -- & $1\times10^{8}$ & $3\times10^{-4}$& \cite{Harman2015, Hu2019} \\
\textbf{CO} & \textbf{X} & \textbf{0} & \textbf{--} & \textbf{variable} & \textbf{$1\times10^{-7}$}  & \\ 
O$_2$ & X & 0 & -- & 0 & 0 & \cite{Hu2012} \\
CH$_4$ & X & 0 & -- & $3\times10^{8}$ & 0 & \cite{Hu2012} \\
NH$_3$ & X & 0 & -- & 0 & 0 & \cite{Harman2015} \\
\cline{2-5}\\
\multicolumn{6}{c}{O$_2$ Runaway in N$_2$-CO$_2$ Atmosphere}\\
\cline{2-5}\\
N$_2$ & C & 0 & 0.99 & -- & -- & \\
CO$_2$ & X & 0 & 0.01 & -- & -- & \\
H$_2$ & X & Diff.-limit. & -- & $3.\times10^{10}$ & 0  &\cite{Hu2012} \\
NO & X & 0 & -- & $1\times10^{8}$ & $3\times10^{-4}$& \cite{Harman2015, Hu2019} \\
CO & X & 0 & -- & $1\times10^{8}$ & $1\times10^{-8}$ & \cite{Kharecha2005} \\ 
\textbf{O$_2$} & \textbf{X} & \textbf{0} & \textbf{--} & \textbf{variable} & \textbf{$1\times10^{-7}$} &   -- \\
CH$_4$ & X & 0 & -- & $3\times10^{8}$ & 0 & \cite{Hu2012} \\
NH$_3$ & X & 0 & -- & 0 & 0 &\cite{Harman2015} \\\cline{2-5}\\
\multicolumn{6}{c}{CH$_4$ Runaway in CO$_2$-N$_2$ Atmosphere}\\
\cline{2-5}\\
N$_2$ & C & 0 & 0.1 & -- & -- & \\
CO$_2$ & X & 0 & 0.9 & -- & -- &  \\
H$_2$ & X & Diff.-limit. & -- & $3.\times10^{10}$ & 0 &\cite{Hu2012} \\
NO & X & 0 & -- & $1\times10^{8}$ & $3\times10^{-4}$& \cite{Harman2015, Hu2019} \\
CO & X & 0 & -- & $1\times10^{8}$ & $1\times10^{-8}$ & \cite{Kharecha2005, Hu2012} \\ 
O$_2$ & X & 0 & -- & 0 & 0 & \cite{Hu2012} \\
\textbf{CH$_4$} & \textbf{X} & \textbf{0} & -- & \textbf{variable} & \textbf{$1\times10^{-7}$} &  \\
NH$_3$ & X & 0 & -- & 0 & 0 &\cite{Harman2015} \\
\cline{2-5}\\
\enddata
\tablenotetext{a}{``X": the full continuity-diffusion equation is solved for the species; ``A": aerosol, settles out of the atmosphere; ``C": chemically inert. Our code also has the capability to designate species as type``F," i.e.  treated as being in photochemical equilibrium \citep{Hu2012}. However, we do not consider any type ``F" species in this work to better enforce mass balance \citep{James2018}.}
\tablenotetext{b}{Escape flux from the TOA; a negative number corresponds to inflow.}
\tablenotetext{c}{Surface mixing ratio relative to dry air}
\tablenotetext{d}{Surface emission flux}
\tablenotetext{e}{Dry deposition velocity, standard case}
\tablecomments{For the bottom boundary condition, either the surface mixing ratio or the surface flux and deposition velocity are specified.} 
\end{deluxetable}
\end{longrotatetable}

\subsection{Results}
\begin{figure}[h]
\centering
\includegraphics[width=\linewidth]{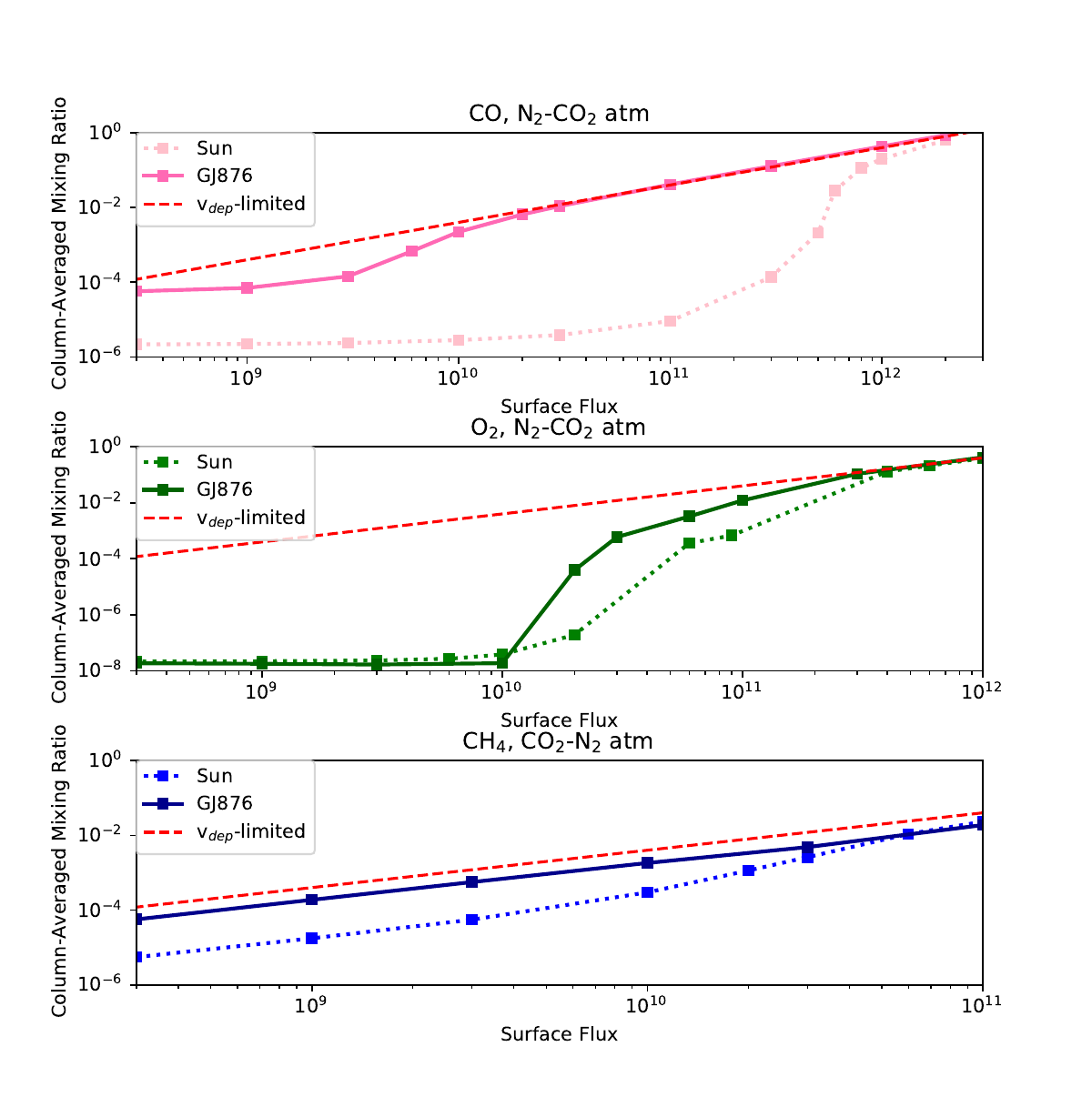}
\caption{Column-averaged CO and O$_2$  (N$_2$-CO$_2$ atmosphere) and CH$_4$ (CO$_2$-N$_2$ atmosphere) mixing ratios as a function of surface emission flux. We reproduce the rapid, nonlinear increases in the concentrations of these gases as a function of surface flux reported or alluded to elsewhere in the literature. Runaway occurs for diverse gases in diverse atmospheres, and is not unique to our model.}
\label{fig:solar_gj876_runaway}
\end{figure}

In this section, we discuss our simulations of CO, O$_2$, and CH$_4$ runaway in the context of the literature (Figure~\ref{fig:solar_gj876_runaway}). Photochemical runaway occurs for a range of gases and atmosphere types, and is not unique to our model.

CO runaway is the best understood due to its photochemical simplicity: the only significant sink of CO is reaction with OH, whose main source in anoxic atmospheres is H$_2$O photolysis. At low $\phi_{CO}$, the concentration of CO, [CO], is a weak function of its surface emission flux $\phi_{CO}$ due to the production of CO by CO$_2$ photolysis; $\phi_{CO}$ must exceed photolytic production to affect [CO]. If $\phi_{CO}$ exceeds the H$_2$O photolysis rate $J_{H_{2}O}$, CO saturates this sink and accumulates rapidly until it is limited by surface deposition \citep{Kasting1990, Kasting2014}. For the temperate, ocean-bearing planetary scenario we consider, $J_{H_{2}O}$ is limited by stellar NUV irradiation; consequently, CO runaway occurs at much lower fluxes on M dwarf planets due to their lower NUV irradiation and hence lower $J_{H_{2}O}$ \citep{Segura2005}. 

O$_2$ runaway is also discussed in the literature, thought not by name. O$_2$ runaway is more complex and challenging to model compared to CO runaway. At low $\phi_{O_{2}}$, [O$_2$] is a weak function of $\phi_{O_{2}}$, due to strong photochemical production in the upper atmosphere from recombination of O derived from CO$_2$ photolysis. At low $\phi_{O_{2}}$, [O$_2$] is limited  by reactions with atmospheric reductants, mediated by UV photons. If $\phi_{O_{2}}$ exceeds the reductant flux into the atmosphere (here, primarily volcanogenic H$_2$, $1.5\times10^{10}$ cm$^2$ s$^{-1}$ O$_2$ equivalents), O$_2$ overcomes its photochemical sinks, and accumulates rapidly as a function of $\phi_{O_{2}}$ until it is limited by surface processes. This is consistent with our understanding of the processes governing the rise of O$_2$ in Earth's atmosphere \citep{Goldblatt2006, Zahnle2006, Daines2017, Gregory2021}. Because the main limit to O$_2$ accumulation is reductant supply and not UV flux, it is not predicted to be significantly easier to accumulate on M dwarf planets, though this is not the case for a more reducing atmosphere (e.g., H$_2$-dominated) where the reductant supply is not limiting.

CH$_4$ runaway has been briefly considered in the literature \citep{Prather1996, Pavlov2003, Segura2005}. For planets orbiting Sun-like stars, OH is the main sink of CH$_4$ at low $\phi_{CH{4}}$. As with CO, when $\phi_{CH{4}}$ grows large enough to complete with $J_{H_{2}O}$, it saturates the OH supply and increases quadratically until it is limited photochemically by direct photolysis. This is consistent with the similarly quadratic growth reported by \citet{Pavlov2003}. CH$_4$ does not enter true runaway (complete removal of all photochemical constraints on buildup) over the range of CH$_4$ emission fluxes we simulate here. For sufficiently high $\phi_{CH{4}}$ we expect that the direct photolysis sink would eventually saturate, but are unable to probe this regime because it would cause [CH$_4$]/[CO$_2$]$>0.1$. Our model is not valid in this regime because our model does not include hydrocarbon haze, and optically thick hydrocarbon haze formation is expected in this regime \citep{DeWitt2009, Arney2016, Arney2017, Hu2012}. Nevertheless, CH$_4$ photolysis is inefficient enough that CH$_4$ still accumulates to very high concentrations \citep{Mount1977, Chen2004, Hu2012}. The situation is different for the GJ 876 case. The low NUV output of our proxy M dwarf means that our CO$_2$-dominated atmosphere destabilizes to CO and O$_2$ \citep{Hu2020}. In this case, CH$_4$ accumulates in an essentially OH-free atmosphere thanks to suppression by high atmospheric CO, and is controlled by photolysis and deposition at all $\phi_{CH{4}}$ modeled here. \citet{Hu2020} report this destabilization to be extremely sensitive to the details of the photochemical network. As a sensitivity test, we prescribed an unrealistically high $v_{dep, O_{2}} = v_{dep, CO} = 1$ cm/s, to force the atmosphere into a low-CO/O$_2$ state. In this case, we again observed a transition from OH attack to direct photolysis as the main limit on [CH$_4$], though OH attack remained a smaller sink than in the solar case because of the lower OH production on M dwarf planets.

\section{Thermodynamic Limits on Runaway\label{sec:microbe-thermo}}
In this section, we quantitatively examine the qualitative argument of \citet{Segura2005} and \citet{Rugheimer2015mdwarf} that runaway is not physical due to biological feedbacks. Specifically, \citet{Segura2005} and \citet{Rugheimer2015mdwarf} argue that microbial methanogenesis would become thermodynamically "unprofitable" at high pCH$_4$, thereby providing a feedback loop to stabilize pCH$_4$ by reducing CH$_4$ production. \citet{Segura2005} and \citet{Rugheimer2015mdwarf} postulate an upper limit on pCH$_4$ of $\sim0.001$ bar. By analogy \citet{Rugheimer2015mdwarf} dismiss the possibility of runaway of other gases, such as N$_2$O, and \citet{Rugheimer2018} dismiss the possibility of runaway for early Earth as well. 

We examine this argument on thermodynamic grounds using the framework of \citet{Seager2013}, who in turn implement the criteria of \citet{Hoehler2004}. We affirm the argument of \citet{Segura2005} and \citet{Rugheimer2015mdwarf} that thermodynamic constraints may prevent methanogenic microbes from triggering a CH$_4$ runaway on modern-Earth-like planets. However, high CH$_4$ (pCH$_4=0.01$ bar) remains thermodynamically compatible with methanogenesis on early-Earth-like worlds, which feature elevated concentrations of the CO$_2$ and H$_2$ that are the substrates of methanogenesis. We analyze the hypothetical "ammoniagenesis" metabolism \citep{Seager2013, Seager2013b, Bains2014}, and show it to be compatible with runaway concentrations of NH$_3$ (pNH$_3=0.001$ bar). Our analysis is simple, and does not encode a full ecosystem model, e.g., \citet{Sauterey2020}. However, this simplicity also reduces the terracentricity of our analysis \citep{Seager2015}, and we contend that it is sufficient to argue against a blanket dismissal of photochemical runaway on purely thermodynamic grounds.

\subsection{Thermodynamics of Metabolic Gas Production\label{sec:thermo-methods}}
We follow \citet{Seager2013} in considering thermodynamic limits on metabolic gas production.

The free energy yielded by a metabolic reaction $\Delta G_{reac}$ can be calculated from thermodynamics, via
$$\Delta G_{reac} = \Delta G^0_{reac} + RT \ln{Q}$$,
where $\Delta G^0_{reac}$ is the standard free energy of the reaction, $T$ is the temperature, $R=8.314$ J mol$^{-1}$ K$^{-1}$ is the universal gas constant, and $Q$ is the reaction quotient.

$\Delta G^0_{reac}$ can be estimated from the difference between the total standard Gibbs free energy of formation $\Delta G^0$ of the products and the reactants, i.e. 
$$\Delta G^0_{reac}=(\Sigma_in_i\Delta G^0_{i}) - (\Sigma_jn_j\Delta G^0_{j})$$,
where $\Delta G^0_{x})$ is the standard free energy of formation of species $x$, $n_x$ is the stoichiometric coefficient associated with species $x$, and the indices $i$ and $j$ count over products and reactants, respectively.

$Q$ is calculated as
$$Q=\frac{\Pi_iA_i^{n_i}}{\Pi_jA_j^{n_j}}$$,
where $A_x$ is the activity of species $x$. We approximate our metabolic reactions as proceeding in dilute aqueous solution, permitting us to ignore salinity effects and to simplify $A_{H_{2}O}=1$. 

For example, for methanogenesis,

\begin{eqnarray}
    CO_2 + 4H_2 \rightarrow CH_4 + 2H_2O\\
    \Delta G^0_{reac}=(\Delta G^0_{CH_{4}}+2\Delta G^0_{H_{2}O})-(\Delta G^0_{CO_{2}}+4\Delta G^0_{H_{2}})\\
    Q=\frac{A_{CH_{4}}A_{H_{2}O}^2}{A_{CO_{2}}A_{H_{2}}^4}=\frac{A_{CH_{4}}}{A_{CO_{2}}A_{H_{2}}^4}
\end{eqnarray}

 We note that $\Delta G^0_{x}$ is a relative quantity, and that the reference species can differ between thermodynamic databases. Therefore, it is important to use an internally consistent thermodynamic database when conducting thermodynamic calculations. We take our $\Delta G^0_{x}(T)$ from \citet{Amend2001}. Further, care must be taken to use the $\Delta G^0_{x}(T)$ corresponding to the phase being considered, i.e. when considering aqueous-phase $x$ a different $\Delta G^0_{x}(T)$ must be used than when considering gas-phase $x$. We do so. 

We can translate $\Delta G_{reac}$ into thermodynamic constraints on microbial metabolic gas production. We consider two constraints, using the terminology of \citet{Hoehler2004}:
\begin{enumerate}
    \item Biological Energy Quantum (BEQ): the Gibbs free energy liberated by metabolism ($\Delta G_{reac}$) must exceed a minimum threshold to be usefully harnessed by the cell. The threshold Gibbs free energy of metabolism required to be usefully harnessed by the cell is empirically constrained for actively growing terrestrial life to be \citep{Hoehler2004}: $$\Delta G_{reac}\leq-20 \texttt{kJ}~\texttt{mol}^{-1}$$. Here, the mol$^{-1}$ refers to mol of reaction, with the reaction written to minimize the number of molecules. We note this constraint to be conservative; one species identified in \citet{Hoehler2004} could survive at $\Delta G_{reac}\leq-5 \texttt{kJ}~\texttt{mol}^{-1}$, suggesting it could grow at $\Delta G_{reac}\leq-10 \texttt{kJ}~\texttt{mol}^{-1}$.
    \item Maintenance Energy: organisms must generate energy at a minimum rate in order survive as active organisms able to grow. This minimum rate increases with temperature. For anaerobic terrestrial life, this minimum energy production rate $P_{me}$ has been empirically measured to be $P_{me}=(2.2\times10^{7} \texttt{kJ}~\texttt{g}^{-1} \texttt{s}^{-1})\exp[\frac{6.94\times10^{4}~\texttt{J}\texttt{mol}^{-1}}{RT}]$ where the g$^{-1}$ refers to grams of wet biomass \citep{Tijhuis1993, Hoehler2004, Seager2013}. $P_{me}$ can be related to $\Delta G_{reac}$ via $P_{me}=\Delta G_{reac}M$, where $M$ is the production rate of the metabolic byproduct in units of mol g$^{-1}$ s$^{-1}$. In the studies we reviewed, $M\leq 10^{-4}$ mol g$^{-1}$ s$^{-1}$ \citep{Muller1986, Patel1978, Patel1977, Pennings2000, Perski1981, Schonheit1985, Takai2008, Zinder1984} and we adopt this as our upper limit, providing our second constraint:
    $$\frac{P_{me}}{\Delta G_{reac}}<10^{-4}~\texttt{mol}~\texttt{g}^{-1}~\texttt{s}^{-1}$$
\end{enumerate}

Next, we apply these constraints to methanogenesis, considered by \citet{Segura2005} and \citet{Rugheimer2015mdwarf}, and to the hypothetical "ammoniagenesis" metabolism postulated in the Cold Haber World scenario we model in this paper. 

\subsection{Thermodynamic Limits on Methanogenesis}
We consider methanogenesis, with general reaction represented by \citep{Bains2014}: $$CO_2 + 4H_2\rightarrow CH_4 + 2H_2O$$

\begin{figure}[h]
\centering
\includegraphics[width=\linewidth]{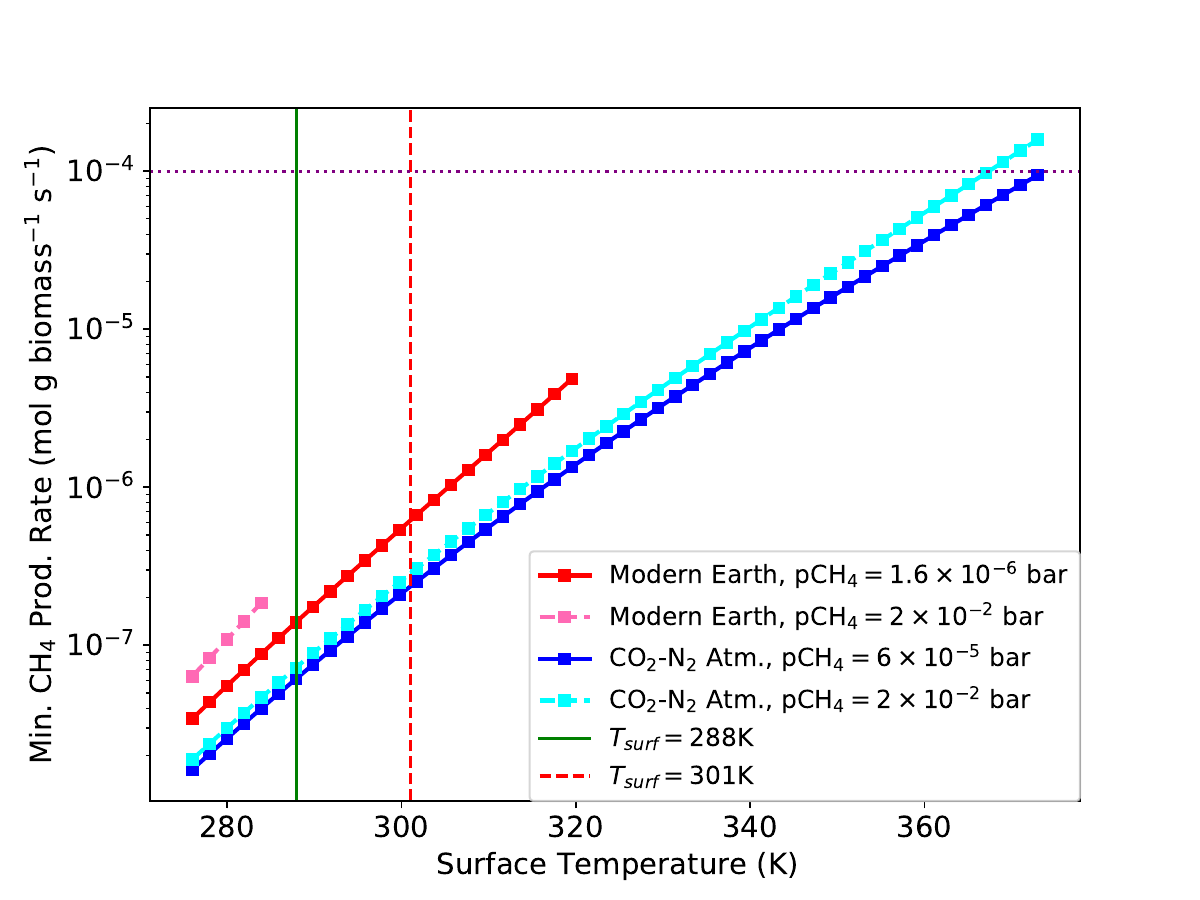}
\caption{Required CH$_4$ production flux for methanogenic life, according to the criteria articulated in Appendix~\ref{sec:microbe-thermo}. The plot shows two constraints. First, the production flux curves terminate at temperatures past which the metabolic reaction no longer yields enough energy to satisfy the BEQ under the given reactant and product concentrations. The pre-runaway $T_{surf}=288$ K is demarcated with a vertical green line, whereas the $T_{surf}\leq301$ K expected in CH$_4$ runaway is demarcated with a vertical red dashed line. Second, the dotted purple line demarcates the empirical limit on the metabolic byproduct production rate observed from modern biology (10$^{-4}$ mol g biomass$^{-1}$ s$^{-1}$). The region above the purple curve violates this empirical constraint, and we consider methanogenesis implausible there. Methanogenesis is thermodynamically unprofitable in CH$_4$ runaway for a modern Earth-like world, but it remains thermodynamically profitable up to very high temperatures for an early Earth-like world, including the including the $T_{surf}\leq$ 301 K expected in CH$_4$ runaway}
\label{fig:methanogenesis}
\end{figure}

Methanogens are strict anaerobes. Consequently, on oxic modern Earth, methanogens are confined to anaerobic environments, with [CO$_2$] present at mM concentrations and [H$_2$] present at nM concentrations \citep{Megonigal2014}. We calculate the $T$ at which methanogenesis satisfies the criteria in Section~\ref{sec:thermo-methods} for [CO$_2$]=3 mM, [H$_2$]=3 nM, and pCH$_4=1.6\times10^{-6}$ bar and $2\times10^{-2}$ bar, corresponding to modern Earth and an Earth in CH$_4$ runaway, respectively (Figure~\ref{fig:methanogenesis}). \text{For modern Earth in CH$_4$ runaway,} we find that methanogenesis is only profitable for $T\leq278$K. For comparison, we calculate $T_{surf}=301$K for this planetary scenario, affirming the argument of \citet{Segura2005} and \citet{Rugheimer2015mdwarf} that thermodynamic feedbacks may inhibit CH$_4$ runaway on a modern-Earth-like planet. However, methanogenesis would be allowed if [CO$_2$] or [H$_2$] were $3\times$higher (e.g., 9 mM, 9 nM), rather than the intermediate values (3 nM, 3 nM) we have chosen here; consequently, we cannot fully rule out the possibility of CH$_4$ runaway on modern Earth-like planets. 

On the other hand, planets more similar to early Earth, i.e. anoxic, with CO$_2$-N$_2$ atmospheres, are more thermodynamically favorable for methanogenesis. We simulate such a planet in Appendix~\ref{sec:runaway_eg}, which features pCO$_2$=0.9 bar, pH$_2=1\times10^{-3}$ bar, and pCH$_4$=$6\times10^{-5}$ and pCH$_4$=$2\times10^{-2}$, corresponding to $\phi_{CH_{4}}=3\times10^{8}$ cm$^{-2}$s$^{-1}$ and $\phi_{CH_{4}}=1\times10^{11}$ cm$^{-2}$s$^{-1}$, respectively. Using the climate model described in Section~\ref{sec:climate_methods}, we predict surface warming of $\leq13$ K corresponding to this CH$_4$ inventory, relative to a base climate with $T_{surf}=288$ K. Methanogenesis remains profitable up to very high temperatures for an early Earth-like world, including the $T_{surf}\leq301$ K (vertical red dashed line) expected in CH$_4$ runaway. 

Note that for the modern Earth case, we considered [CO$_2$] and [H$_2$], whereas for the early Earth-like case, we considered pCO$_2$ and pH$_2$. For these calculations, we have been careful to use $\Delta G^0_{x}(T)$ from \cite{Amend2001} corresponding to the aqueous phase and gas phase, respectively. 

\subsection{Thermodynamic Limits on Ammonia Production in Cold Haber World Scenario}
We consider the hypothetical "ammoniagenesis" metabolism proposed for Cold Haber Worlds and modeled in this paper, with the general reaction \citep{Seager2013, Seager2013b, Bains2014}
$$ N_2 + 3H_2\rightarrow 2NH_3$$

\begin{figure}[h]
\centering
\includegraphics[width=\linewidth]{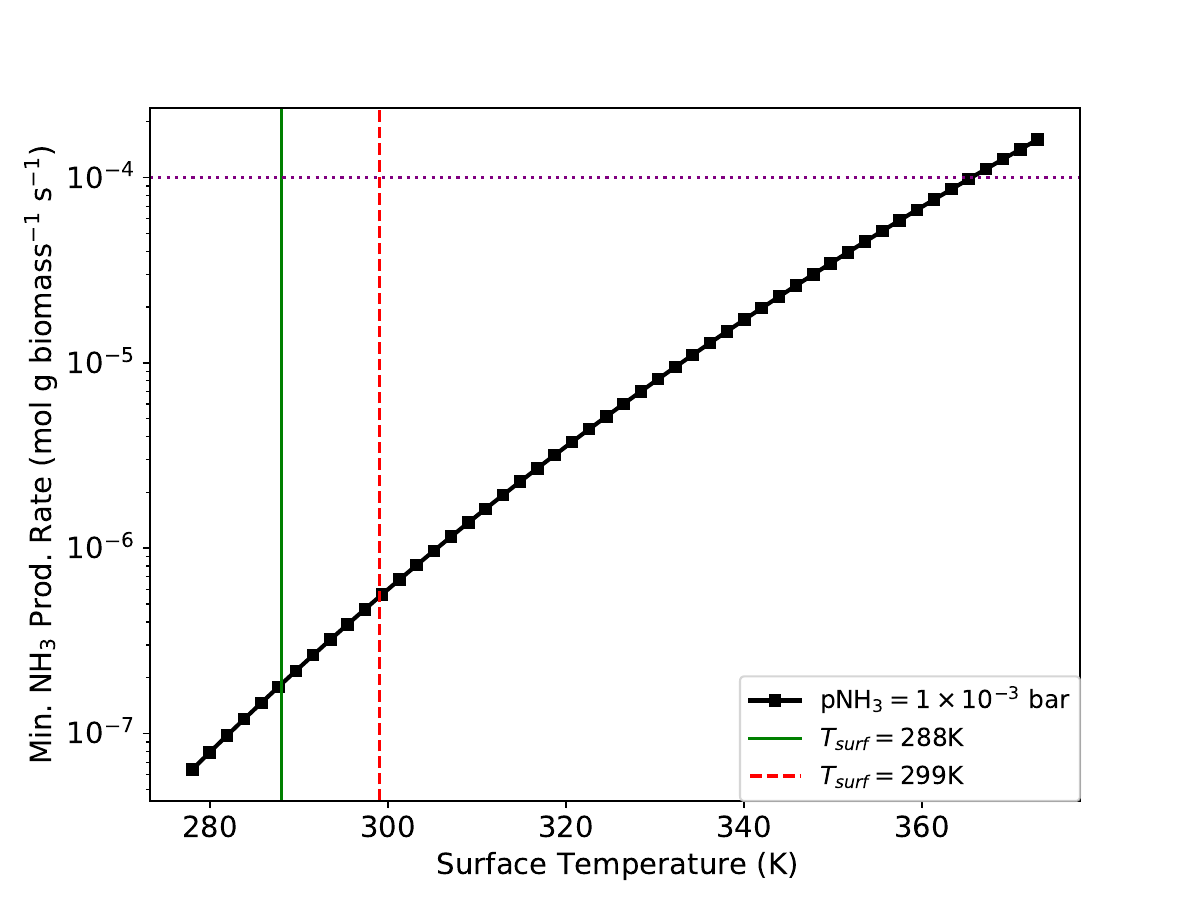}
\caption{Required NH$_3$ production flux for life metabolizing using the hypothetical ammoniagenesis pathway, according to the criteria articulated in Appendix~\ref{sec:microbe-thermo}. The plot shows two constraints. First, the production flux curves terminate at temperatures past which the metabolic reaction no longer yields enough energy to satisfy the BEQ under the given reactant and product concentrations. The pre-runaway $T_{surf}=288$ K is demarcated with a vertical green line, whereas the $T_{surf}=299$ K expected in NH$_3$ runaway is demarcated with the vertical dashed red line. Second, the dotted purple line demarcates the empirical limit on the metabolic byproduct production rate observed from modern biology (10$^{-4}$ mol g biomass$^{-1}$ s$^{-1}$). The region above the purple curve violates this empirical constraint, and we consider ammoniagenesis implausible there. Ammoniagenesis remains profitable up to very high temperatures for a Cold Haber World, including the $T_{surf}$ = 299 K expected in NH$_3$ runaway.}
\label{fig:ammoniagenesis}
\end{figure}

For the atmospheric scenario we simulate in this paper, pH$_2$=0.9 bar, pN$_2$=0.1 bar, and pNH$_3$=0.001 bar in runaway. For these parameters, the ammoniagenesis is profitable for $T\leq360$ K (Fig.~\ref{fig:ammoniagenesis}). For comparison, we predict $T_{surf}=299$K for pNH$_3=7.6\times10^{-4}$ bar. We conclude that ammoniagenesis remains profitable in NH$_3$ runaway, based on the criteria from \citet{Hoehler2004}.

We further considered whether an H$_2$-dominated atmosphere is \emph{required} for the Cold Haber World. We reran the calculation above for pN$_2$=0.1 bar and pH$_2$=0.01 bar, and found that ammoniagenesis remains thermodynamically profitable across a wide range of habitable temperatures. Early Earth-like planets are predicted to host H$_2$ mixing ratios of up to 10\% if H$_2$ escape is below the diffusion limit, due to, e.g., cooling of the upper atmosphere by high CO$_2$ abundances \citep{Tian2005, Liggins2020}. This implies that even an early Earth-like planet with a non-H$_2$ dominated atmosphere would be thermodynamically compatible with ammoniagenesis, though the metabolic incentive to develop this metabolism would be proportionately reduced.

\section{Failure Modes for Photochemical Runaway\label{sec:runaway_fail}}
In this section, we consider pathological scenarios in which gases might fail to enter photochemical runaway despite high production flux. 

The key requirement for photochemical runaway is that the radical production in the atmosphere vary sublinearly with the runaway gas flux, so that the runaway gas can saturate its photochemical sinks. One scenario in which this condition fails corresponds to combustion-explosion conditions. In conditions of sufficiently extreme disequilibrium (e.g.,  the coexistence of sufficiently-abundant H$_2$ and O$_2$), chain propagation reactions (radicals tend to make more radicals) can dominate over chain termination reactions (radicals tend to make nonradicals). In such conditions, energetic events can trigger runaway increase in radical concentrations until the substrates are exhausted. This condition can be checked by comparing to experimental combustion-explosion studies \citep{Grenfell2018}. 

More generally, there may exist cases in which a biosignature gas self-limits its concentrations, e.g. through self-reaction or through the production of UV sensitizer. Photochemical models can check for this possibility, but careful study is required. For example, O$_2$ does produce a UV sensitizer in the form of O$_3$, but far from catalyzing its destruction, O$_3$ photochemically shields O$_2$ \citep{Zahnle2006}. Detailed photochemical networks are not available for many potential biosignature gases; the measurement of these photochemical data and their incorporation into photochemical models is required to explore the potential of novel biosignature gases for runaway.


\bibliography{sample631}{}

\begin{thebibliography}{}
\expandafter\ifx\csname natexlab\endcsname\relax\def\natexlab#1{#1}\fi
\providecommand{\url}[1]{\href{#1}{#1}}
\providecommand{\dodoi}[1]{doi:~\href{http://doi.org/#1}{\nolinkurl{#1}}}
\providecommand{\doeprint}[1]{\href{http://ascl.net/#1}{\nolinkurl{http://ascl.net/#1}}}
\providecommand{\doarXiv}[1]{\href{https://arxiv.org/abs/#1}{\nolinkurl{https://arxiv.org/abs/#1}}}

\bibitem[{Amend \& Shock(2001)}]{Amend2001}
Amend, J.~P., \& Shock, E.~L. 2001, FEMS microbiology reviews, 25, 175

\bibitem[{{Arney} {et~al.}(2018){Arney}, {Domagal-Goldman}, \&
  {Meadows}}]{Arney2018}
{Arney}, G., {Domagal-Goldman}, S.~D., \& {Meadows}, V.~S. 2018, Astrobiology,
  18, 311, \dodoi{10.1089/ast.2017.1666}

\bibitem[{{Arney} {et~al.}(2016){Arney}, {Domagal-Goldman}, {Meadows}, {Wolf},
  {Schwieterman}, {Charnay}, {Claire}, {H{\'e}brard}, \& {Trainer}}]{Arney2016}
{Arney}, G., {Domagal-Goldman}, S.~D., {Meadows}, V.~S., {et~al.} 2016,
  Astrobiology, 16, 873, \dodoi{10.1089/ast.2015.1422}

\bibitem[{{Arney} {et~al.}(2017){Arney}, {Meadows}, {Domagal-Goldman},
  {Deming}, {Robinson}, {Tovar}, {Wolf}, \& {Schwieterman}}]{Arney2017}
{Arney}, G.~N., {Meadows}, V.~S., {Domagal-Goldman}, S.~D., {et~al.} 2017,
  \apj, 836, 49, \dodoi{10.3847/1538-4357/836/1/49}

\bibitem[{{Atreya} {et~al.}(1977){Atreya}, {Donahue}, \& {Kuhn}}]{Atreya1977}
{Atreya}, S.~K., {Donahue}, T.~M., \& {Kuhn}, W.~R. 1977, \icarus, 31, 348,
  \dodoi{10.1016/0019-1035(77)90027-6}

\bibitem[{Bains {et~al.}(2014)Bains, Seager, \& Zsom}]{Bains2014}
Bains, W., Seager, S., \& Zsom, A. 2014, Life, 4, 716

\bibitem[{{Batalha} {et~al.}(2018){Batalha}, {Lewis}, {Line}, {Valenti}, \&
  {Stevenson}}]{Batalha2018}
{Batalha}, N.~E., {Lewis}, N.~K., {Line}, M.~R., {Valenti}, J., \& {Stevenson},
  K. 2018, \apjl, 856, L34, \dodoi{10.3847/2041-8213/aab896}

\bibitem[{{Batalha} {et~al.}(2017){Batalha}, {Mandell}, {Pontoppidan},
  {Stevenson}, {Lewis}, {Kalirai}, {Earl}, {Greene}, {Albert}, \&
  {Nielsen}}]{Batalha2017}
{Batalha}, N.~E., {Mandell}, A., {Pontoppidan}, K., {et~al.} 2017, \pasp, 129,
  064501, \dodoi{10.1088/1538-3873/aa65b0}

\bibitem[{Bouwman {et~al.}(1997)Bouwman, Lee, Asman, Dentener, Van Der~Hoek, \&
  Olivier}]{Bouwman1997}
Bouwman, A., Lee, D., Asman, W., {et~al.} 1997, Global biogeochemical cycles,
  11, 561

\bibitem[{Catling \& Kasting(2017)}]{CatlingKasting2017}
Catling, D.~C., \& Kasting, J.~F. 2017, Atmospheric evolution on inhabited and
  lifeless worlds (Cambridge University Press)

\bibitem[{{Chen} \& {Wu}(2004)}]{Chen2004}
{Chen}, F.~Z., \& {Wu}, C.~Y.~R. 2004, \jqsrt, 85, 195,
  \dodoi{10.1016/S0022-4073(03)00225-5}

\bibitem[{Cohen \& Westberg(1991)}]{Cohen1991}
Cohen, N., \& Westberg, K. 1991, Journal of physical and chemical reference
  data, 20, 1211

\bibitem[{{Correia} {et~al.}(2010){Correia}, {Couetdic}, {Laskar}, {Bonfils},
  {Mayor}, {Bertaux}, {Bouchy}, {Delfosse}, {Forveille}, {Lovis}, {Pepe},
  {Perrier}, {Queloz}, \& {Udry}}]{Correia2010}
{Correia}, A.~C.~M., {Couetdic}, J., {Laskar}, J., {et~al.} 2010, \aap, 511,
  A21, \dodoi{10.1051/0004-6361/200912700}

\bibitem[{{Cowan} {et~al.}(2015){Cowan}, {Greene}, {Angerhausen}, {Batalha},
  {Clampin}, {Col{\'o}n}, {Crossfield}, {Fortney}, {Gaudi}, {Harrington},
  {Iro}, {Lillie}, {Linsky}, {Lopez-Morales}, {Mandell}, \&
  {Stevenson}}]{Cowan2015}
{Cowan}, N.~B., {Greene}, T., {Angerhausen}, D., {et~al.} 2015, \pasp, 127,
  311, \dodoi{10.1086/680855}

\bibitem[{{Curdt} {et~al.}(2004){Curdt}, {Landi}, \& {Feldman}}]{Curdt2004}
{Curdt}, W., {Landi}, E., \& {Feldman}, U. 2004, \aap, 427, 1045,
  \dodoi{10.1051/0004-6361:20041278}

\bibitem[{{Daines} {et~al.}(2017){Daines}, {Mills}, \& {Lenton}}]{Daines2017}
{Daines}, S.~J., {Mills}, B. J.~W., \& {Lenton}, T.~M. 2017, Nature
  Communications, 8, 14379, \dodoi{10.1038/ncomms14379}

\bibitem[{{Deming} \& {Seager}(2017)}]{Deming2017}
{Deming}, L.~D., \& {Seager}, S. 2017, Journal of Geophysical Research
  (Planets), 122, 53, \dodoi{10.1002/2016JE005155}

\bibitem[{{DeWitt} {et~al.}(2009){DeWitt}, {Trainer}, {Pavlov}, {Hasenkopf},
  {Aiken}, {Jimenez}, {McKay}, {Toon}, \& {Tolbert}}]{DeWitt2009}
{DeWitt}, H.~L., {Trainer}, M.~G., {Pavlov}, A.~A., {et~al.} 2009,
  Astrobiology, 9, 447, \dodoi{10.1089/ast.2008.0289}

\bibitem[{{Domagal-Goldman} {et~al.}(2011){Domagal-Goldman}, {Meadows},
  {Claire}, \& {Kasting}}]{Domagal-Goldman2011}
{Domagal-Goldman}, S.~D., {Meadows}, V.~S., {Claire}, M.~W., \& {Kasting},
  J.~F. 2011, Astrobiology, 11, 419, \dodoi{10.1089/ast.2010.0509}

\bibitem[{{Fauchez} {et~al.}(2020){Fauchez}, {Villanueva}, {Schwieterman},
  {Turbet}, {Arney}, {Pidhorodetska}, {Kopparapu}, {Mandell}, \&
  {Domagal-Goldman}}]{Fauchez2020}
{Fauchez}, T.~J., {Villanueva}, G.~L., {Schwieterman}, E.~W., {et~al.} 2020,
  Nature Astronomy, 4, 372, \dodoi{10.1038/s41550-019-0977-7}

\bibitem[{{France} {et~al.}(2016){France}, {Parke Loyd}, {Youngblood}, {Brown},
  {Schneider}, {Hawley}, {Froning}, {Linsky}, {Roberge}, {Buccino},
  {Davenport}, {Fontenla}, {Kaltenegger}, {Kowalski}, {Mauas}, {Miguel},
  {Redfield}, {Rugheimer}, {Tian}, {Vieytes}, {Walkowicz}, \&
  {Weisenburger}}]{France2016}
{France}, K., {Parke Loyd}, R.~O., {Youngblood}, A., {et~al.} 2016,
  Astrophysical Journal, 820, 89, \dodoi{10.3847/0004-637X/820/2/89}

\bibitem[{{Gardner} {et~al.}(2006){Gardner}, {Mather}, {Clampin}, {Doyon},
  {Greenhouse}, {Hammel}, {Hutchings}, {Jakobsen}, {Lilly}, {Long}, {Lunine},
  {McCaughrean}, {Mountain}, {Nella}, {Rieke}, {Rieke}, {Rix}, {Smith},
  {Sonneborn}, {Stiavelli}, {Stockman}, {Windhorst}, \& {Wright}}]{Gardner2006}
{Gardner}, J.~P., {Mather}, J.~C., {Clampin}, M., {et~al.} 2006, \ssr, 123,
  485, \dodoi{10.1007/s11214-006-8315-7}

\bibitem[{{Gebauer} {et~al.}(2017){Gebauer}, {Grenfell}, {Stock}, {Lehmann},
  {Godolt}, {von Paris}, \& {Rauer}}]{Gebauer2017}
{Gebauer}, S., {Grenfell}, J.~L., {Stock}, J.~W., {et~al.} 2017, Astrobiology,
  17, 27, \dodoi{10.1089/ast.2015.1384}

\bibitem[{{Goldblatt} {et~al.}(2006){Goldblatt}, {Lenton}, \&
  {Watson}}]{Goldblatt2006}
{Goldblatt}, C., {Lenton}, T.~M., \& {Watson}, A.~J. 2006, \nat, 443, 683,
  \dodoi{10.1038/nature05169}

\bibitem[{{Gordon} {et~al.}(2017){Gordon}, {Rothman}, {Hill}, {Kochanov},
  {Tan}, {Bernath}, {Birk}, {Boudon}, {Campargue}, {Chance}, {Drouin}, {Flaud},
  {Gamache}, {Hodges}, {Jacquemart}, {Perevalov}, {Perrin}, {Shine}, {Smith},
  {Tennyson}, {Toon}, {Tran}, {Tyuterev}, {Barbe}, {Cs{\'a}sz{\'a}r}, {Devi},
  {Furtenbacher}, {Harrison}, {Hartmann}, {Jolly}, {Johnson}, {Karman},
  {Kleiner}, {Kyuberis}, {Loos}, {Lyulin}, {Massie}, {Mikhailenko},
  {Moazzen-Ahmadi}, {M{\"u}ller}, {Naumenko}, {Nikitin}, {Polyansky}, {Rey},
  {Rotger}, {Sharpe}, {Sung}, {Starikova}, {Tashkun}, {Auwera}, {Wagner},
  {Wilzewski}, {Wcis{\l}o}, {Yu}, \& {Zak}}]{Gordon2017}
{Gordon}, I.~E., {Rothman}, L.~S., {Hill}, C., {et~al.} 2017, \jqsrt, 203, 3,
  \dodoi{10.1016/j.jqsrt.2017.06.038}

\bibitem[{Goumri {et~al.}(1999)Goumri, Rocha, Laakso, Smith, \&
  Marshall}]{Goumri1999}
Goumri, A., Rocha, J.-D.~R., Laakso, D., Smith, C., \& Marshall, P. 1999, The
  Journal of Physical Chemistry A, 103, 11328

\bibitem[{{Gregory} {et~al.}(2021){Gregory}, {Claire}, \&
  {Rugheimer}}]{Gregory2021}
{Gregory}, B.~S., {Claire}, M.~W., \& {Rugheimer}, S. 2021, Earth and Planetary
  Science Letters, 561, 116818, \dodoi{10.1016/j.epsl.2021.116818}

\bibitem[{{Grenfell} {et~al.}(2018){Grenfell}, {Gebauer}, {Godolt}, {Stracke},
  {Lehmann}, \& {Rauer}}]{Grenfell2018}
{Grenfell}, J.~L., {Gebauer}, S., {Godolt}, M., {et~al.} 2018, \apj, 861, 38,
  \dodoi{10.3847/1538-4357/aab2a9}

\bibitem[{{Harman} {et~al.}(2018){Harman}, {Felton}, {Hu}, {Domagal-Goldman},
  {Segura}, {Tian}, \& {Kasting}}]{Harman2018}
{Harman}, C.~E., {Felton}, R., {Hu}, R., {et~al.} 2018, \apj, 866, 56,
  \dodoi{10.3847/1538-4357/aadd9b}

\bibitem[{{Harman} {et~al.}(2015){Harman}, {Schwieterman}, {Schottelkotte}, \&
  {Kasting}}]{Harman2015}
{Harman}, C.~E., {Schwieterman}, E.~W., {Schottelkotte}, J.~C., \& {Kasting},
  J.~F. 2015, Astrophysical Journal, 812, 137,
  \dodoi{10.1088/0004-637X/812/2/137}

\bibitem[{Hoehler(2004)}]{Hoehler2004}
Hoehler, T. 2004, Geobiology, 2, 205

\bibitem[{Hu \& Diaz(2019)}]{Hu2019}
Hu, R., \& Diaz, H.~D. 2019, The Astrophysical Journal, 886, 126

\bibitem[{{Hu} {et~al.}(2020){Hu}, {Peterson}, \& {Wolf}}]{Hu2020}
{Hu}, R., {Peterson}, L., \& {Wolf}, E.~T. 2020, \apj, 888, 122,
  \dodoi{10.3847/1538-4357/ab5f07}

\bibitem[{{Hu} {et~al.}(2012){Hu}, {Seager}, \& {Bains}}]{Hu2012}
{Hu}, R., {Seager}, S., \& {Bains}, W. 2012, Astrophysical Journal, 761, 166,
  \dodoi{10.1088/0004-637X/761/2/166}

\bibitem[{{Hu} {et~al.}(2013){Hu}, {Seager}, \& {Bains}}]{Hu2013}
---. 2013, Astrophysical Journal, 769, 6, \dodoi{10.1088/0004-637X/769/1/6}

\bibitem[{{Huang} {et~al.}(2022){Huang}, {Seager}, {Petkowski}, {Ranjan}, \&
  {Zhan}}]{Huang2020}
{Huang}, J., {Seager}, S., {Petkowski}, J.~J., {Ranjan}, S., \& {Zhan}, Z.
  2022, Astrobiology, 22, 171, \dodoi{10.1089/ast.2020.2358}

\bibitem[{Jacob(1999)}]{Jacob1999}
Jacob, D.~J. 1999, Introduction to atmospheric chemistry (Princeton University
  Press)

\bibitem[{{James} \& {Hu}(2018)}]{James2018}
{James}, T., \& {Hu}, R. 2018, \apj, 867, 17, \dodoi{10.3847/1538-4357/aae2bb}

\bibitem[{Kaltenegger(2017)}]{Kaltenegger2017}
Kaltenegger, L. 2017, Annual Review of Astronomy and Astrophysics, 55, 433

\bibitem[{{Kasting}(1982)}]{Kasting1982}
{Kasting}, J.~F. 1982, Journal of Geophysics Research, 87, 3091,
  \dodoi{10.1029/JC087iC04p03091}

\bibitem[{{Kasting}(1990)}]{Kasting1990}
---. 1990, Origins of Life and Evolution of the Biosphere, 20, 199,
  \dodoi{10.1007/BF01808105}

\bibitem[{Kasting(2014)}]{Kasting2014}
Kasting, J.~F. 2014, Geological Society of America Special Papers, 504, 19

\bibitem[{{Kasting} {et~al.}(2014){Kasting}, {Kopparapu}, {Ramirez}, \&
  {Harman}}]{Kasting2014PNAS}
{Kasting}, J.~F., {Kopparapu}, R., {Ramirez}, R.~M., \& {Harman}, C.~E. 2014,
  Proceedings of the National Academy of Science, 111, 12641,
  \dodoi{10.1073/pnas.1309107110}

\bibitem[{{Kawashima} \& {Rugheimer}(2019)}]{Kawashima2019}
{Kawashima}, Y., \& {Rugheimer}, S. 2019, \aj, 157, 213,
  \dodoi{10.3847/1538-3881/ab14e3}

\bibitem[{Kharecha {et~al.}(2005)Kharecha, Kasting, \& Siefert}]{Kharecha2005}
Kharecha, P., Kasting, J., \& Siefert, J. 2005, Geobiology, 3, 53

\bibitem[{{Koll} \& {Cronin}(2019)}]{KollCronin2019}
{Koll}, D. D.~B., \& {Cronin}, T.~W. 2019, \apj, 881, 120,
  \dodoi{10.3847/1538-4357/ab30c4}

\bibitem[{{Kozakis} {et~al.}(2020){Kozakis}, {Lin}, \&
  {Kaltenegger}}]{Kozakis2020}
{Kozakis}, T., {Lin}, Z., \& {Kaltenegger}, L. 2020, \apjl, 894, L6,
  \dodoi{10.3847/2041-8213/ab6f6a}

\bibitem[{{Kuhn} \& {Atreya}(1979)}]{Kuhn1979}
{Kuhn}, W.~R., \& {Atreya}, S.~K. 1979, \icarus, 37, 207,
  \dodoi{10.1016/0019-1035(79)90126-X}

\bibitem[{Laufer \& Fahr(2004)}]{Laufer2004}
Laufer, A.~H., \& Fahr, A. 2004, Chemical reviews, 104, 2813

\bibitem[{{Liggins} {et~al.}(2020){Liggins}, {Shorttle}, \&
  {Rimmer}}]{Liggins2020}
{Liggins}, P., {Shorttle}, O., \& {Rimmer}, P.~B. 2020, Earth and Planetary
  Science Letters, 550, 116546, \dodoi{10.1016/j.epsl.2020.116546}

\bibitem[{{Lin} {et~al.}(2022){Lin}, {Seager}, {Ranjan}, {Kozakis}, \&
  {Kaltenegger}}]{Lin2022}
{Lin}, Z., {Seager}, S., {Ranjan}, S., {Kozakis}, T., \& {Kaltenegger}, L.
  2022, \apjl, 925, L10, \dodoi{10.3847/2041-8213/ac4788}

\bibitem[{{Loyd} {et~al.}(2016){Loyd}, {France}, {Youngblood}, {Schneider},
  {Brown}, {Hu}, {Linsky}, {Froning}, {Redfield}, {Rugheimer}, \&
  {Tian}}]{Loyd2016}
{Loyd}, R.~O.~P., {France}, K., {Youngblood}, A., {et~al.} 2016, Astrophysical
  Journal, 824, 102, \dodoi{10.3847/0004-637X/824/2/102}

\bibitem[{{Lustig-Yaeger} {et~al.}(2019){Lustig-Yaeger}, {Meadows}, \&
  {Lincowski}}]{Lustig-Yaeger2019}
{Lustig-Yaeger}, J., {Meadows}, V.~S., \& {Lincowski}, A.~P. 2019, \aj, 158,
  27, \dodoi{10.3847/1538-3881/ab21e0}

\bibitem[{Lyons {et~al.}(2014)Lyons, Reinhard, \& Planavsky}]{Lyons2014}
Lyons, T.~W., Reinhard, C.~T., \& Planavsky, N.~J. 2014, Nature, 506, 307

\bibitem[{{Madhusudhan} {et~al.}(2021){Madhusudhan}, {Piette}, \&
  {Constantinou}}]{Madhusudhan2021}
{Madhusudhan}, N., {Piette}, A. A.~A., \& {Constantinou}, S. 2021, \apj, 918,
  1, \dodoi{10.3847/1538-4357/abfd9c}

\bibitem[{Meadows {et~al.}(2018)Meadows, Reinhard, Arney, Parenteau,
  Schwieterman, Domagal-Goldman, Lincowski, Stapelfeldt, Rauer, DasSarma,
  {et~al.}}]{Meadows2018}
Meadows, V.~S., Reinhard, C.~T., Arney, G.~N., {et~al.} 2018, Astrobiology, 18,
  630

\bibitem[{Megonigal {et~al.}(2014)Megonigal, Hines, \&
  Visscher}]{Megonigal2014}
Megonigal, J., Hines, M., \& Visscher, P. 2014, in Treatise on Geochemistry
  (Second Edition), second edition edn., ed. H.~D. Holland \& K.~K. Turekian
  (Oxford: Elsevier), 273--359,
  \dodoi{https://doi.org/10.1016/B978-0-08-095975-7.00808-1}

\bibitem[{M{\"o}ller \& Wagner(1984)}]{Moller1984}
M{\"o}ller, W., \& Wagner, H.~G. 1984, Zeitschrift f{\"u}r Naturforschung A,
  39, 846

\bibitem[{{Morley} {et~al.}(2017){Morley}, {Kreidberg}, {Rustamkulov},
  {Robinson}, \& {Fortney}}]{Morley2017}
{Morley}, C.~V., {Kreidberg}, L., {Rustamkulov}, Z., {Robinson}, T., \&
  {Fortney}, J.~J. 2017, \apj, 850, 121, \dodoi{10.3847/1538-4357/aa927b}

\bibitem[{{Mount} {et~al.}(1977){Mount}, {Warden}, \& {Moos}}]{Mount1977}
{Mount}, G.~H., {Warden}, E.~S., \& {Moos}, H.~W. 1977, Astrophysical Journall,
  214, L47, \dodoi{10.1086/182440}

\bibitem[{M{\"u}ller {et~al.}(1986)M{\"u}ller, Blaut, \&
  Gottschalk}]{Muller1986}
M{\"u}ller, V., Blaut, M., \& Gottschalk, G. 1986, Applied and environmental
  microbiology, 52, 269

\bibitem[{{Owen} {et~al.}(2020){Owen}, {Shaikhislamov}, {Lammer}, {Fossati}, \&
  {Khodachenko}}]{Owen2020}
{Owen}, J.~E., {Shaikhislamov}, I.~F., {Lammer}, H., {Fossati}, L., \&
  {Khodachenko}, M.~L. 2020, \ssr, 216, 129, \dodoi{10.1007/s11214-020-00756-w}

\bibitem[{Patel {et~al.}(1978)Patel, Khan, \& Roth}]{Patel1978}
Patel, G., Khan, A., \& Roth, L. 1978, Journal of Applied Bacteriology, 45, 347

\bibitem[{Patel \& Roth(1977)}]{Patel1977}
Patel, G., \& Roth, L. 1977, Canadian journal of microbiology, 23, 893

\bibitem[{{Pavlov} {et~al.}(2003){Pavlov}, {Hurtgen}, {Kasting}, \&
  {Arthur}}]{Pavlov2003}
{Pavlov}, A.~A., {Hurtgen}, M.~T., {Kasting}, J.~F., \& {Arthur}, M.~A. 2003,
  Geology, 31, 87, \dodoi{10.1130/0091-7613(2003)031<0087:MRPA>2.0.CO;2}

\bibitem[{Pennings {et~al.}(2000)Pennings, Vermeij, de~Poorter, Keltjens, \&
  Vogels}]{Pennings2000}
Pennings, J.~L., Vermeij, P., de~Poorter, L.~M., Keltjens, J.~T., \& Vogels,
  G.~D. 2000, Antonie van Leeuwenhoek, 77, 281

\bibitem[{Perski {et~al.}(1981)Perski, Moll, \& Thauer}]{Perski1981}
Perski, H.-J., Moll, J., \& Thauer, R.~K. 1981, Archives of Microbiology, 130,
  319

\bibitem[{{Phillips} {et~al.}(2021){Phillips}, {Wang}, {Kendrew}, {Greene},
  {Hu}, {Valenti}, {Panero}, \& {Schulze}}]{Phillips2021}
{Phillips}, C.~L., {Wang}, J., {Kendrew}, S., {et~al.} 2021, \apj, 923, 144,
  \dodoi{10.3847/1538-4357/ac29be}

\bibitem[{{Prather}(1996)}]{Prather1996}
{Prather}, M.~J. 1996, \grl, 23, 2597, \dodoi{10.1029/96GL02371}

\bibitem[{{Ranjan} {et~al.}(2020){Ranjan}, {Schwieterman}, {Harman}, {Fateev},
  {Sousa-Silva}, {Seager}, \& {Hu}}]{Ranjan2020}
{Ranjan}, S., {Schwieterman}, E.~W., {Harman}, C., {et~al.} 2020, \apj, 896,
  148, \dodoi{10.3847/1538-4357/ab9363}

\bibitem[{{Reinhard} {et~al.}(2017){Reinhard}, {Olson}, {Schwieterman}, \&
  {Lyons}}]{Reinhard2017}
{Reinhard}, C.~T., {Olson}, S.~L., {Schwieterman}, E.~W., \& {Lyons}, T.~W.
  2017, Astrobiology, 17, 287, \dodoi{10.1089/ast.2016.1598}

\bibitem[{{Rodler} \& {L{\'o}pez-Morales}(2014)}]{Rodler2014}
{Rodler}, F., \& {L{\'o}pez-Morales}, M. 2014, Astrophysical Journal, 781, 54,
  \dodoi{10.1088/0004-637X/781/1/54}

\bibitem[{{Rogers}(2015)}]{Rogers2015}
{Rogers}, L.~A. 2015, Astrophysical Journal, 801, 41,
  \dodoi{10.1088/0004-637X/801/1/41}

\bibitem[{Rugheimer \& Kaltenegger(2018)}]{Rugheimer2018}
Rugheimer, S., \& Kaltenegger, L. 2018, The Astrophysical Journal, 854, 19

\bibitem[{Rugheimer {et~al.}(2015)Rugheimer, Kaltenegger, Segura, Linsky, \&
  Mohanty}]{Rugheimer2015mdwarf}
Rugheimer, S., Kaltenegger, L., Segura, A., Linsky, J., \& Mohanty, S. 2015,
  The Astrophysical Journal, 809, 57

\bibitem[{{Sauterey} {et~al.}(2020){Sauterey}, {Charnay}, {Affholder},
  {Mazevet}, \& {Ferri{\`e}re}}]{Sauterey2020}
{Sauterey}, B., {Charnay}, B., {Affholder}, A., {Mazevet}, S., \&
  {Ferri{\`e}re}, R. 2020, Nature Communications, 11, 2705,
  \dodoi{10.1038/s41467-020-16374-7}

\bibitem[{Sch{\"o}nheit \& Beimborn(1985)}]{Schonheit1985}
Sch{\"o}nheit, P., \& Beimborn, D.~B. 1985, European journal of biochemistry,
  148, 545

\bibitem[{{Schwieterman} {et~al.}(2019){Schwieterman}, {Reinhard}, {Olson},
  {Ozaki}, {Harman}, {Hong}, \& {Lyons}}]{Schwieterman2019}
{Schwieterman}, E.~W., {Reinhard}, C.~T., {Olson}, S.~L., {et~al.} 2019, \apj,
  874, 9, \dodoi{10.3847/1538-4357/ab05e1}

\bibitem[{Seager \& Bains(2015)}]{Seager2015}
Seager, S., \& Bains, W. 2015, Science advances, 1, e1500047

\bibitem[{{Seager} {et~al.}(2013{\natexlab{a}}){Seager}, {Bains}, \&
  {Hu}}]{Seager2013}
{Seager}, S., {Bains}, W., \& {Hu}, R. 2013{\natexlab{a}}, Astrophysical
  Journal, 775, 104, \dodoi{10.1088/0004-637X/775/2/104}

\bibitem[{{Seager} {et~al.}(2013{\natexlab{b}}){Seager}, {Bains}, \&
  {Hu}}]{Seager2013b}
---. 2013{\natexlab{b}}, \apj, 777, 95, \dodoi{10.1088/0004-637X/777/2/95}

\bibitem[{{Seager} {et~al.}(2020){Seager}, {Huang}, {Petkowski}, \&
  {Pajusalu}}]{Seager2020}
{Seager}, S., {Huang}, J., {Petkowski}, J.~J., \& {Pajusalu}, M. 2020, Nature
  Astronomy, 4, 802, \dodoi{10.1038/s41550-020-1069-4}

\bibitem[{{Segura} {et~al.}(2005){Segura}, {Kasting}, {Meadows}, {Cohen},
  {Scalo}, {Crisp}, {Butler}, \& {Tinetti}}]{Segura2005}
{Segura}, A., {Kasting}, J.~F., {Meadows}, V., {et~al.} 2005, Astrobiology, 5,
  706, \dodoi{10.1089/ast.2005.5.706}

\bibitem[{Seinfeld \& Pandis(2016)}]{SeinfeldPandis}
Seinfeld, J.~H., \& Pandis, S.~N. 2016, Atmospheric chemistry and physics: from
  air pollution to climate change (John Wiley \& Sons)

\bibitem[{Siddique {et~al.}(2017)Siddique, Altarawneh, Gore, Westmoreland, \&
  Dlugogorski}]{Siddique2017}
Siddique, K., Altarawneh, M., Gore, J., Westmoreland, P.~R., \& Dlugogorski,
  B.~Z. 2017, The Journal of Physical Chemistry A, 121, 2221

\bibitem[{{Sousa-Silva} {et~al.}(2020){Sousa-Silva}, {Seager}, {Ranjan},
  {Petkowski}, {Zhan}, {Hu}, \& {Bains}}]{Sousa-Silva2020}
{Sousa-Silva}, C., {Seager}, S., {Ranjan}, S., {et~al.} 2020, Astrobiology, 20,
  235, \dodoi{10.1089/ast.2018.1954}

\bibitem[{{Stassun} {et~al.}(2019){Stassun}, {Oelkers}, {Paegert}, {Torres},
  {Pepper}, {De Lee}, {Collins}, {Latham}, {Muirhead}, {Chittidi},
  {Rojas-Ayala}, {Fleming}, {Rose}, {Tenenbaum}, {Ting}, {Kane}, {Barclay},
  {Bean}, {Brassuer}, {Charbonneau}, {Ge}, {Lissauer}, {Mann}, {McLean},
  {Mullally}, {Narita}, {Plavchan}, {Ricker}, {Sasselov}, {Seager}, {Sharma},
  {Shiao}, {Sozzetti}, {Stello}, {Vanderspek}, {Wallace}, \&
  {Winn}}]{Stassun2019}
{Stassun}, K.~G., {Oelkers}, R.~J., {Paegert}, M., {et~al.} 2019, \aj, 158,
  138, \dodoi{10.3847/1538-3881/ab3467}

\bibitem[{Takai {et~al.}(2008)Takai, Nakamura, Toki, Tsunogai, Miyazaki,
  Miyazaki, Hirayama, Nakagawa, Nunoura, \& Horikoshi}]{Takai2008}
Takai, K., Nakamura, K., Toki, T., {et~al.} 2008, Proceedings of the National
  Academy of Sciences, 105, 10949

\bibitem[{{Tian} {et~al.}(2011){Tian}, {Kasting}, \& {Zahnle}}]{Tian2011}
{Tian}, F., {Kasting}, J.~F., \& {Zahnle}, K. 2011, Earth and Planetary Science
  Letters, 308, 417, \dodoi{10.1016/j.epsl.2011.06.011}

\bibitem[{{Tian} {et~al.}(2005){Tian}, {Toon}, {Pavlov}, \& {De
  Sterck}}]{Tian2005}
{Tian}, F., {Toon}, O.~B., {Pavlov}, A.~A., \& {De Sterck}, H. 2005, Science,
  308, 1014, \dodoi{10.1126/science.1106983}

\bibitem[{Tijhuis {et~al.}(1993)Tijhuis, Van~Loosdrecht, \&
  Heijnen}]{Tijhuis1993}
Tijhuis, L., Van~Loosdrecht, M.~C., \& Heijnen, J.~v. 1993, Biotechnology and
  bioengineering, 42, 509

\bibitem[{{Wordsworth} {et~al.}(2017){Wordsworth}, {Kalugina}, {Lokshtanov},
  {Vigasin}, {Ehlmann}, {Head}, {Sanders}, \& {Wang}}]{Wordsworth2017}
{Wordsworth}, R., {Kalugina}, Y., {Lokshtanov}, S., {et~al.} 2017, Geophysical
  Research Letters, 44, 665, \dodoi{10.1002/2016GL071766}

\bibitem[{{Youngblood} {et~al.}(2016){Youngblood}, {France}, {Loyd}, {Linsky},
  {Redfield}, {Schneider}, {Wood}, {Brown}, {Froning}, {Miguel}, {Rugheimer},
  \& {Walkowicz}}]{Youngblood2016}
{Youngblood}, A., {France}, K., {Loyd}, R.~O.~P., {et~al.} 2016, \apj, 824,
  101, \dodoi{10.3847/0004-637X/824/2/101}

\bibitem[{Zahnle {et~al.}(2006)Zahnle, Claire, \& Catling}]{Zahnle2006}
Zahnle, K., Claire, M., \& Catling, D. 2006, Geobiology, 4, 271

\bibitem[{{Zahnle}(1986)}]{Zahnle1986}
{Zahnle}, K.~J. 1986, Journal of Geophysics Research, 91, 2819,
  \dodoi{10.1029/JD091iD02p02819}

\bibitem[{{Zeng} {et~al.}(2019){Zeng}, {Jacobsen}, {Sasselov}, {Petaev},
  {Vanderburg}, {Lopez-Morales}, {Perez-Mercader}, {Mattsson}, {Li}, {Heising},
  {Bonomo}, {Damasso}, {Berger}, {Cao}, {Levi}, \& {Wordsworth}}]{Zeng2019}
{Zeng}, L., {Jacobsen}, S.~B., {Sasselov}, D.~D., {et~al.} 2019, Proceedings of
  the National Academy of Science, 116, 9723, \dodoi{10.1073/pnas.1812905116}

\bibitem[{{Zhan} {et~al.}(2021){Zhan}, {Seager}, {Petkowski}, {Sousa-Silva},
  {Ranjan}, {Huang}, \& {Bains}}]{Zhan2020}
{Zhan}, Z., {Seager}, S., {Petkowski}, J.~J., {et~al.} 2021, Astrobiology, 21,
  765, \dodoi{10.1089/ast.2019.2146}

\bibitem[{{Zheng} {et~al.}(2008){Zheng}, {Jewitt}, {Osamura}, \&
  {Kaiser}}]{Zheng2008}
{Zheng}, W., {Jewitt}, D., {Osamura}, Y., \& {Kaiser}, R.~I. 2008, \apj, 674,
  1242, \dodoi{10.1086/523783}

\bibitem[{Zhu {et~al.}(2015)Zhu, Henze, Bash, Cady-Pereira, Shephard, Luo, \&
  Capps}]{Zhu2015}
Zhu, L., Henze, D.~K., Bash, J.~O., {et~al.} 2015, Current Pollution Reports,
  1, 95

\bibitem[{Zinder \& Koch(1984)}]{Zinder1984}
Zinder, S.~H., \& Koch, M. 1984, Archives of Microbiology, 138, 263

\end{thebibliography}
\bibliographystyle{aasjournal}



\end{document}